\documentclass[usenames,dvipsnames]{article}
\usepackage[utf8]{inputenc}
\usepackage[a4paper,top=2.5cm,bottom=2.5cm,left=2cm,right=2cm]{geometry}
\usepackage[english]{babel}
\usepackage[justification=centering]{caption}
\usepackage[square, comma, sort&compress, numbers]{natbib}
\usepackage{lmodern}
\usepackage{titling} 
\usepackage{floatflt,epsfig}
\usepackage{tabularx}
\usepackage{amsmath}
\usepackage{amsfonts}
\usepackage{float}
\usepackage{multirow}
\usepackage{wrapfig}
\usepackage{amssymb}
\usepackage{graphicx}
\usepackage{caption}
\usepackage{booktabs}
\usepackage{siunitx}
\usepackage{amssymb,amsmath,amsthm,amsfonts} 
\usepackage[italian]{varioref} 
\usepackage{braket}
\usepackage{comment}
\usepackage{subcaption}
\usepackage{slashed}
\usepackage{cancel}
\usepackage{verbatim}
\numberwithin{equation}{section}
\usepackage{feynmp-auto,expdlist}
\usepackage{latexsym}
\usepackage{hepnicenames}
\usepackage{enumerate}
\usepackage{soul}
\usepackage[normalem]{ulem}
\usepackage{wasysym}
\usepackage{makecell}
\usepackage{bbm}
\usepackage[compat=1.1.0]{tikz-feynman}
\usepackage{hyperref}
\usepackage{xcolor}

\begin{document}

\begin{center}  
{\LARGE
\bf
The Chiral Lagrangian of CP-Violating \\ 
\smallskip
Axion-Like Particles
} \\
\vspace{0.8cm}

{\bf Luca Di Luzio$^{a}$, Gabriele Levati$^{a,b}$, Paride Paradisi$^{a,b}$}\\[7mm]

{\it $^{a}$Istituto Nazionale di Fisica Nucleare (INFN), Sezione di Padova, \\
Via F. Marzolo 8, 35131 Padova, Italy}\\[1mm]
{\it $^{b}$Dipartimento di Fisica e Astronomia `G.~Galilei', Universit\`a di Padova,
 \\ Via F. Marzolo 8, 35131 Padova, Italy
}\\[1mm]

\vspace{0.3cm}

\begin{quote}
We discuss the construction of the most general CP-violating chiral Lagrangian for an axion-like particle (ALP). Starting with an effective Lagrangian containing light quarks and gluons, we provide its matching onto a chiral effective Lagrangian 
at $\mathcal{O}(p^2)$ described in terms of mesons 
and baryons, identifying the correspondence between the Jarlskog invariants of the two theories. After deriving the ALP interactions with mesons and baryons, we analyse a few
relevant phenomenological implications such as the permanent electric dipole moments of nucleons and the CP-violating 
ALP and kaon decays.
This work provides the necessary tools for further phenomenological analyses connecting low-energy observables with the couplings of the underlying ultraviolet complete theory.

\end{quote}

\thispagestyle{empty}
\end{center}

\bigskip

\tableofcontents

\clearpage



\section{Introduction}


The lack for any convincing evidence of heavy new physics (NP) at the LHC has motivated an increasing theoretical as well as experimental investigation of scenarios with new light mediators.
The most popular example is provided by extensions of the Standard Model (SM) entailing light pseudoscalar bosons, generically referred to as axion-like particles (ALPs)~\cite{Jaeckel:2010ni}. 
The emergence of light ALPs can be naturally justified if they are pseudo-Nambu-Goldstone bosons associated with the spontaneous breaking of an underlying global symmetry.
Remarkably, ALPs can address several open problems in particle physics like the strong 
CP problem~\cite{Peccei:1977ur,Peccei:1977hh,Weinberg:1977ma,Wilczek:1977pj}, the evidence of dark matter~\cite{Preskill:1982cy,Abbott:1982af,Dine:1982ah}, 
as well as the flavour~\cite{Davidson:1981zd,Wilczek:1982rv} and 
the weak-scale
hierarchy~\cite{Graham:2015cka} problems. 

Standard QCD axion models~\cite{Kim:1979if,Shifman:1979if,Zhitnitsky:1980tq,Dine:1981rt}, originally invoked as a solution to the strong CP problem, predict a strict relation between the axion mass and decay constant. 
However, since this relation depends on the specific ultraviolet (UV) completion, 
a model-independent approach is commonly assumed where ALPs are treated as a generalization of the QCD axion, with unrelated mass and couplings to be probed experimentally. ALP interactions with SM fermions and gauge bosons are then described 
via an effective Lagrangian containing operators up to dimension-5~\cite{Georgi:1986df}. 
This approach provides the opportunity to look for light ALPs with masses in the range between an MeV and tens of GeV, whose couplings are not tightly constrained either by cosmological and astrophysical bounds~\cite{Marsh:2015xka,Irastorza:2018dyq,DiLuzio:2020wdo} or by collider searches~\cite{Mimasu:2014nea,Jaeckel:2015jla,Knapen:2016moh,Brivio:2017ije,Bauer:2017nlg,Bauer:2017ris,Mariotti:2017vtv,Bauer:2018uxu,Aloni:2018vki,Alonso-Alvarez:2018irt,Aloni:2019ruo,Gavela:2019cmq}.
Moreover, low-energy observables that are very sensitive probes of ALP interactions are given by flavor-changing neutral-currents (FCNC) processes, both in the quark~\cite{Batell:2009jf,Gavela:2019wzg,MartinCamalich:2020dfe,Bauer:2021mvw,DiLuzio:2023ndz,Cornella:2023kjq}
and in the lepton sectors~\cite{Bauer:2019gfk,Cornella:2019uxs,Calibbi:2020jvd}. Indeed, since there is no fundamental motivation for ALP interactions to obey the SM flavor group, ALPs can induce FCNC already at the tree level.

So far, most of the phenomenological analyses related to ALPs were inherent to CP conserving phenomena (see, however, Refs.~\cite{Marciano:2016yhf,Stadnik:2017hpa,Bertolini:2020hjc,DiLuzio:2020oah,Gorghetto:2021luj,Okawa:2021fto,Dekens:2022gha,Bonnefoy:2022rik,DasBakshi:2023lca}).
The CP symmetry is violated provided ALP couplings to photons (or equivalently to gluons)
include both $\phi F\tilde F$ and $\phi F F$ interactions (where $F$ is the QED field strength tensor, $\tilde F$ its dual and $\phi$ denotes the ALP field) and/or if ALP couplings to fermions ($f$) account for both $\phi \bar f i\gamma_5 f$ and $\phi \bar f f$ interactions. 
These requirements are fulfilled if the global shift symmetry, protecting the ALP mass, as well as the CP symmetry are broken 
by the UV sector of the theory and such a symmetry breaking is eventually transmitted to the infrared sector by some dynamical mechanism. 
The necessary UV dynamics can naturally arise in strongly-coupled theories, and in fact an explicit realization arises in the case of the neutral pion in the SM through QCD and electromagnetic interactions (see e.g.~\cite{Choi:1990cn}). 
A strongly-interacting sector emerging at a scale $\Lambda \gtrsim 1$ TeV which resembles the QCD dynamics is therefore  conceivable.

A CP-violating ALP is also motivated by relaxion models \cite{Graham:2015cka} addressing the weak-scale hierarchy problem by introducing an ALP field, the relaxion, which scans the Higgs boson mass in the early universe from a scale $M \gg 1$ TeV down to the electroweak scale. In this case, the simultaneous presence of the relaxion-Higgs mixing and the relaxion-photon/gluon couplings induces CP violation.

From the experimental side, there is an outstanding program to improve the current limits of permanent electric dipole moments (EDMs) of nuclei, nucleons, atoms and molecules by few orders of magnitude \cite{Chupp:2017rkp}. 
This motivated the analysis of Ref.~\cite{DiLuzio:2020oah} where the CP-violating Jarlskog invariants emerging in the ALP effective field theory (EFT) were classified and the leading short-distance effects on EDMs up to two loops were evaluated. 
However, Ref.~\cite{DiLuzio:2020oah} assumed ALP masses above the GeV scale where QCD can be treated perturbatively. 

The main goal of the present work is to extend the analysis of Ref.~\cite{DiLuzio:2020oah} to ALP masses in the well-motivated sub-GeV region where non-perturbative methods, such as chiral perturbation theory ($\chi \text{pt}$), have to be employed.
The construction of the effective chiral Lagrangian for an ALP interacting with photons and light pseudoscalar mesons has been 
already discussed at length in the literature for the case of a 
CP-odd ALP (see Refs.~\cite{Georgi:1986df,Bauer:2020jbp,Bauer:2021wjo,Bandyopadhyay:2021wbb}, 
as well as \cite{GrillidiCortona:2015jxo,Vonk:2020zfh,Vonk:2021sit,DiLuzio:2021vjd,DiLuzio:2022tbb} 
for analyses beyond leading order in $\chi \text{pt}$).
In our work, we will generalise the above studies by including both 
CP-odd and CP-even ALP interactions with mesons and baryons.
In particular, we will discuss in detail the matching of the perturbative 
effective Lagrangian at low energies onto a chiral effective Lagrangian,  
identifying the relations between the high-energy Jarlskog invariants
and those emerging in the chiral theory.
The results of this work represent the basis for phenomenological analyses related to a light CP-violating ALP, connecting low-energy observables, such as EDMs and 
flavour-violating processes entailing mesons and baryons, with the couplings of the underlying UV complete theory.

The paper is structured as follows. 
In Section \ref{sec:ChiPT_For_ALPs}, we present the construction of the most general 
CP-violating chiral Lagrangian for ALPs describing 
interactions 
with mesons and baryons. In Section \ref{sec:ALPINT2F}, we apply the general
results of Section \ref{sec:ChiPT_For_ALPs} to a two-flavour setting, therefore obtaining 
explicit expressions for ALP interactions with pions and nucleons.
Section \ref{sec:Nf3} is instead relative to a three-flavour setup where explicit ALP interactions with pions, kaons, and the eta meson as well as with baryons will be presented. Section \ref{sec:pheno} is dedicated to a short phenomenological analysis including the predictions for the proton and neutron EDMs and CP-violating ALP and kaons decays.
Conclusions and an outlook for future possible studies, which can be obtained starting from our results, are presented in Section \ref{sec:concl}.

\section{Construction of the CP-violating Chiral Lagrangian for ALPs}
\label{sec:ChiPT_For_ALPs}

In this section, we are going to build the most general chiral Lagrangian describing 
CP-violating interactions of a light ALP with mesons and baryons.
 We will consider ALP interactions with SM fields up to order $\mathcal{O}(1/\Lambda)$ in the $\Lambda$ expansion, while we will limit ourselves to the leading order interactions in the chiral expansion. More explicitly, this means that we will only consider interactions of order $\mathcal{O}(p^2)$ between ALPs and mesons and up to order $\mathcal{O}(p)$ between ALPs and baryons.

First, let us consider the $SU(3)_c \times U(1)_{\text{em}}$ invariant dimension-5 effective Lagrangian accounting for CP-violating interactions of an ALP with photons, gluons and light quarks~\cite{DiLuzio:2020oah}: 
\begin{equation}
\label{eq:AAA}
\begin{split}
    \mathcal{L}_{\text{ALP}}^{\text{QCD}} &= 
    \frac{1}{2} \partial_\mu \phi \partial^\mu \phi - \frac{1}{2} M_\phi^2 \phi^2
    + e^2 \frac{\tilde{C}_\gamma}{\Lambda} \, \phi \, F^{\mu\nu} \tilde{F}_{\mu\nu}
    + g_s^2 \frac{\tilde{C}_g}{\Lambda}\, \phi \, G_a^{\mu\nu} \tilde{G}^a_{\mu\nu}
    + \frac{\partial_\mu \phi}{\Lambda} \bar{q} \, \gamma^\mu (Y_S + Y_P \gamma_5) \, q
      \\
    & \quad 
    + e^2 \frac{C_\gamma}{\Lambda} \, \phi \, F^{\mu\nu} F_{\mu\nu}
    + g_s^2 \frac{C_g}{\Lambda} \, \phi \, G_a^{\mu\nu} G^a_{\mu\nu} 
 + \frac{v}{\Lambda} \, \phi \,  \bar{q} \, y_{S}\,  q  + \mathcal{O} \left(\frac{1}{\Lambda^2}\right) \, ,
\end{split}
\end{equation}
that is valid above the GeV scale. Here, $\phi$ is the ALP field, while 
$\Lambda \gg v \approx 246$ GeV is the EFT cutoff scale, 
where the electroweak vacuum expectation value keeps track of the $SU(2)_L \times U(1)_Y$ origin of the effective operator. 
Moreover,
$q^T = (u,d,s)$ and $Y_S$, $Y_P$ 
and $y_{S}$ are hermitian matrices.  
 $G_{\mu \nu}^a$ and $F_{\mu\nu}$ are the QCD and QED field-strength tensors, while $\tilde{G}_{\mu \nu}^a = 1/2 \,  \varepsilon_{\mu\nu\rho\sigma}G^{\rho \sigma}_a$ and $\tilde{F}_{\mu\nu} = 1/2 \,  \varepsilon_{\mu\nu\rho\sigma} F^{\rho \sigma}$ are their duals. 
Note that the interaction terms in the 
first line of Eq.~(\ref{eq:AAA}) preserve the ALP 
shift symmetry (up to non-perturbative effects in the case of $\phi G \tilde G$), while the mass term, $M_\phi$, 
and the interaction terms in the second line of Eq.~(\ref{eq:AAA}) explicitly break it. See Ref.~\cite{Bonnefoy:2022rik}
for a more detailed 
account of the shift-breaking orders in the ALP effective Lagrangian.
 In particular, a potential term for the ALP involving self-interactions could be added to our Lagrangian. 
Although this terms could in principle 
contribute to phenomenological observables, such as EDMs, 
they can be safely neglected due to their strong $1/\Lambda^n$ suppression, where $n$ is the number of ALP fields appearing in an interaction vertex from such potential terms.

Below the GeV scale, the  derivation of the chiral Lagrangian associated to \eqref{eq:AAA}
is implemented according to the following steps. First of all, it is convenient to remove the gluonic scalar and pseudoscalar densities, $\phi GG$ and $\phi G \tilde{G}$ respectively, in favor of other quantities.
As far as the coupling $\phi GG$ is concerned, we exploit the trace anomaly which allows to trade $\phi GG$ for the trace of the symmetrized QCD energy-momentum tensor $\theta^{\mu\nu}$ \cite{Leutwyler:1989tn}:
\begin{equation}
\label{eq:Trace_Anomaly}
 \theta^\mu_\mu = 
\bar{q} M_q q
 - \frac{ \alpha_s}{8 \pi} \, \beta^0_{\text{QCD}}\,  G^{\mu\nu}_a G_{\mu\nu}^a - \frac{ \alpha_{\text{em}}}{8 \pi}\, \beta^0_{\text{QED}}\,  
 F^{\mu\nu} F_{\mu\nu} \, , 
\end{equation}
 where $M_q = \text{diag} (m_u, m_d, m_s)$ is the quark mass matrix. Moreover, $\beta^0_{\text{QCD}} = (11 N_c - 2 n_f)/3$ and $\beta^0_{\text{QED}} = - 4/3 \sum_{q= \{u,d,s\}} Q_q^2 N_c^q$ are the leading-order beta function coefficients for QCD and QED with three active light quarks, respectively.
As a result, we obtain 
\begin{equation}
\begin{split}
    \mathcal{L}_{\text{ALP}}^{\text{QCD}} &= e^2 \frac{C'_\gamma}{\Lambda} \, \phi \, F^{\mu\nu} F_{\mu\nu} + e^2 \frac{\tilde{C}_\gamma}{\Lambda} \, \phi \, F^{\mu\nu} \tilde{F}_{\mu\nu} + g_s^2 \frac{\tilde{C}_g}{\Lambda}\, \phi \, G_a^{\mu\nu} \tilde{G}^a_{\mu\nu} - \frac{8\pi}{\alpha_s}\frac{g_s^2 C_g}{\beta^0_\text{QCD}}\, \frac{\phi}{\Lambda} \,\theta^\mu_\mu \\
    & \quad + \frac{\partial_\mu \phi}{\Lambda} \bar{q} \, \gamma^\mu (Y_S + Y_P \gamma_5) \, q + \phi\, \frac{v}{\Lambda} \bar{q} \, \mathcal{Z}\,  q +
    \mathcal{O} \left(\frac{1}{\Lambda^2}\right) \, , 
\end{split}
\end{equation}
where
we have: 
\begin{equation}
\begin{split}
    C'_\gamma &= C_\gamma - \frac{\beta^0_{\text{QED}}}{\beta^0_{\text{QCD}}} C_g\,,\\
    \mathcal{Z} &= y_{S}  + \frac{1}{\beta^0_{\text{QCD}}} \frac{8\pi}{\alpha_s} \frac{M_q}{v} g_s^2 C_g\,. 
\end{split}
\end{equation}
Instead, the coupling $\phi G \tilde{G}$ can be removed performing the following ALP-dependent quark redefinition: 
\begin{equation}
\label{eq:Quark_Field_Redefinition}
 q \rightarrow \exp \left[ i \frac{\phi}{\Lambda} \left( Q_V + \lambda_g Q_A \gamma_5 \right) \right] q \, ,
\end{equation}
where $Q_V $ and $Q_A $ are arbitrary $3\times3$ hermitian matrices.
Such a field redefinition is anomalous under both $SU(3)_c$ and $U(1)_{\text{em}}$, therefore inducing the following shifts in the Lagrangian:
\begin{equation}
\begin{split}
    \delta \mathcal{L}_g &= -g_s^2 \,\frac{\phi}{\Lambda} G \tilde{G} \frac{\lambda_{g}}{16\pi^2}\,  \text{tr} (Q_A) \, , \\
    \delta \mathcal{L}_\gamma &= -e^2 \,\frac{\phi}{\Lambda} F \tilde{F} \frac{N_c \lambda_{g}}{8\pi^2}\,  \text{tr} (Q_A Q_q^2) \, , 
\end{split}
\end{equation}
where $Q_q = \text{diag } (2/3, -1/3, -1/3)$ is the charge matrix for light quarks. Requiring that $\lambda_g = 32 \pi^2 \tilde{C}_g$ and $\text{tr} (Q_A) = \frac{1}{2}$, the term $\phi G \tilde{G}$ simply drops out of the Lagrangian while the coupling 
$\tilde{C}_{\gamma}$ is shifted into
\begin{equation}
 \tilde{C}_{\gamma} \rightarrow \tilde{C}_\gamma - 4 N_c \tilde{C}_g \, \text{tr} (Q_A Q_q^2)\,.
\end{equation}
The field redefinition \eqref{eq:Quark_Field_Redefinition} has yet two other consequences. Via the kinetic term of quarks, it modifies the couplings to fermions according to 

\begin{equation}
\begin{split}
Y_S & \rightarrow Y_S - Q_V \, , \\
Y_P & \rightarrow Y_P - \lambda_g Q_A = Y_P - 32 \pi^2 \, \tilde{C}_g Q_A  \, ,
\end{split}
\end{equation}
and it also affects the quark mass term, which gets dressed by the ALP field:
\begin{equation}
M_q \quad  \rightarrow  \quad
e^{i \frac{\phi}{\Lambda}\lambda_gQ_A } M_q e^{i \frac{\phi}{\Lambda}\lambda_g Q_A } 
\equiv M_q^\phi\,.  
\end{equation}
Clearly, also the scalar coupling $\bar{q} \mathcal{Z} q$ and the energy-momentum tensor are affected by such a rotation. However, it is simple to convince oneself that these modification would only generate couplings of two ALPs with quarks that are suppressed by at least two powers of $\Lambda$.

As a result, we end up with the following quark Lagrangian
\begin{equation}
\label{eq:QCD_Scale_ALP_Lag_V1}
\begin{split}
\mathcal{L}_{\text{ALP}}^{\text{QCD}} &= e^2 \frac{c_\gamma}{\Lambda} \, \phi \,  F^{\mu\nu}F_{\mu\nu} + e^2 \frac{\tilde{c}_\gamma}{\Lambda} \, \phi \, F^{\mu\nu}\tilde{F}_{\mu\nu} + \frac{\partial_\mu \phi }{\Lambda}\, \bar{q} \gamma^\mu \left(Y_V + Y_A \gamma_5\right) q - \kappa \, \frac{\phi}{\Lambda} \, \theta^\mu_\mu + \frac{v}{\Lambda} \, \phi \, \bar{q} \mathcal{Z} q  \, + \mathcal{O}\left(\frac{1}{\Lambda^2}\right) \, ,
\end{split}
\end{equation}
where we have defined the following parameters:
\begin{eqnarray}
 c_\gamma  &=& \, C_\gamma - \frac{\beta^0_{\text{QED}}}{\beta^0_{\text{QCD}}} C_g \, ,\\
    \tilde{c}_\gamma &=& \, \tilde{C}_\gamma  -4 N_c \text{ tr }(Q_AQ_q^2)  \tilde{C}_g  \, , \\
    Y^{ij}_V &=& Y^{ij}_{S} - Q^{ij}_V \, , \\
    Y^{ij}_A &=& Y_{P}^{ij} - 32 \pi^2 \, Q_A^{ij} \tilde{C}_g \, , \\
    \mathcal{Z} &=& y_{S}  + \frac{g_s^2 C_g }{\beta^0_{\text{QCD}}} \frac{8\pi}{\alpha_s} \frac{M_q}{v}  =  y_{S}  +\kappa \, \frac{M_q}{v}  \, .
\end{eqnarray}
Finally, making use of the completeness relation $2\, T^a_{ab}T^a_{cd} = \delta_{ad}\delta_{bc} - 1/N \delta_{ab} \delta_{cd}$ valid for the $SU(N)$ group, the quark currents 
can be written as
\begin{equation}
\label{eq:Completeness_Currents}
\begin{split}
    & \bar{q} \, \gamma^\mu Y_V \,q  = 
    2\, \text{Tr} (Y_V T_a) \,  \bar{q} \, T^a \gamma^\mu q + \frac{1}{N} \text{Tr} (Y_V)\, \bar{q}\, \gamma^\mu q  \\
    & \bar{q} \, \gamma^\mu \gamma_5 Y_A\, q = 
        2\, \text{Tr} (Y_A T_a) \,  \bar{q} \, T^a \gamma^\mu \gamma_5 q + \frac{1}{N} \text{Tr} (Y_A)\, \bar{q}\, \gamma^\mu \gamma_5 q
\end{split}
\end{equation}
which in turn modifies \eqref{eq:QCD_Scale_ALP_Lag_V1} into 
\begin{equation}
\label{eq:QCD_Scale_ALP_Lag_V2}
\begin{split}
\mathcal{L}_{\text{ALP}}^{\text{QCD}} &= \frac{\partial_\mu \phi }{\Lambda}\, 
\left[  
2\, \text{Tr} (Y_V T_a) \, j_{V, a}^\mu + 
\frac{1}{N}\, \text{Tr} (Y_V) \, j_{V}^\mu + 
2\, \text{Tr} (Y_A T_a) \, j_{A, a}^\mu + 
\frac{1}{N}\, \text{Tr} (Y_A) \, j_{A}^\mu
\right] \\
&\quad + \frac{v}{\Lambda} \, \phi \, \bar{q} \mathcal{Z} q - \kappa \, \frac{\phi}{\Lambda} \, \theta^\mu_\mu +e^2 \frac{c_\gamma}{\Lambda} \, \phi \,  F^{\mu\nu}F_{\mu\nu} + e^2 \frac{\tilde{c}_\gamma}{\Lambda} \, \phi \, F^{\mu\nu}\tilde{F}_{\mu\nu}  \, ,
\end{split}
\end{equation}
where we have defined the following currents:
\begin{equation}
\label{eq:Completeness_Currents}
\begin{split}
    & j_{V, a}^\mu = \bar{q} \, T^a \gamma^\mu q  \, ,
    \qquad\qquad\qquad j_{V}^\mu = \bar{q} \gamma^\mu q  \, , \\
    &  j_{A, a}^\mu = \bar{q} \, T^a \gamma^\mu \gamma_5\, q  \, , 
     \qquad\qquad\quad  j_{A}^\mu  = \bar{q} \, \gamma^\mu \gamma_5\, q   \,.
\end{split}
\end{equation}
In the following, we will trade quarks for 
hadrons (mesons and baryons) exploiting the 
hadron-quark duality.

\subsection{ALP interactions with Mesons}

At energies above few GeV, ALP interactions with quarks are correctly described by the Lagrangian of Eq. (2.1). 
Instead, for the description of strong interactions at energies below the GeV scale, we can exploit the $\chi \text{pt}$ formalism~\cite{Gasser:1984gg, Pich:1995bw}.
Let us consider first the massless QCD Lagrangian based on the chiral symmetry group $G = SU(3)_L \times SU(3)_R$:
\begin{equation}
\label{eq:Free_QCD_Lag}
\mathcal{L}_{\text{QCD}}^0 = -\frac{1}{4} G^a_{\mu\nu}G_a^{\mu\nu} + i \bar{q}_L \gamma^\mu D_\mu q_L+ i \bar{q}_R \gamma^\mu D_\mu q_R \, , 
\end{equation}
where $q = (u, d, s)^T$.
Chiral symmetry-breaking terms (like mass terms or interactions with external gauge fields other than gluons) can be implemented in $\mathcal{L}_{\text{QCD}}^0$ by introducing appropriate spurions ($a_\mu, v_\mu , s, p$) as external source fields: 
\begin{equation}
\label{eq:QCD_Lag_With_Sources}
\begin{split}
\mathcal{L}_{\text{QCD}}^{\text{ext}} &= \mathcal{L}_{\text{QCD}}^0 + \bar{q} \gamma^\mu (v_\mu + a_\mu \gamma_5)q - \bar{q} (s - i p \gamma_5) q\\
&= \mathcal{L}_{\text{QCD}}^0 + \bar{q} \gamma^\mu (2 r_\mu P_R + 2 \ell_\mu P_L)q - \bar{q} (s - i p \gamma_5) q \,.
\end{split}
\end{equation}
Its chiral counterpart is then found to be 
\begin{equation}
\label{eq:Chipt_Lag_With_Sources}
\mathcal{L}_{\chi \text{pt}}^{\text{ext}}  = \frac{f_\pi^2}{4} \text{Tr}\left[D_\mu \Sigma^\dagger D^\mu \Sigma + \Sigma^\dagger \chi + \chi^\dagger \Sigma\right] +\mathcal{O} (p^4) \, ,
\end{equation}
where $\Sigma (x) = \exp \left[ i \lambda_a \pi_a(x)/f_\pi\right]$ 
is the mesonic matrix transforming as $\Sigma(x) \rightarrow R \Sigma(x) L^\dagger$
under 
$SU(3)_L \times SU(3)_R$, $\pi_a(x)$ are the Goldstone boson fields of $SU(3)_L \times SU(3)_R \to SU(3)_V$ spontaneous symmetry breaking and $f_\pi = 92.4 \pm 0.3$ MeV is the pion decay constant. Moreover, we defined
\begin{equation}
 D_\mu \Sigma = \partial_\mu \Sigma - i r_\mu \, \Sigma + i \Sigma \, \ell_\mu  \qquad \text{and} \qquad \chi = 2B_0 \, (s + i p)\,.
\end{equation}
The duality existing between the Lagrangians \eqref{eq:QCD_Lag_With_Sources} and \eqref{eq:Chipt_Lag_With_Sources} can be established via the path-integral description of the generating functional $Z\left[v_\mu, a_\mu , s, p\right]$:
\begin{equation}
\label{eq:QCD_Chipt_Duality}
 \exp (i Z) = \int \mathcal{D}q\, \mathcal{D} \bar{q}\, \mathcal{D}G_\mu\, \exp \left( i \int d^4 x \,\mathcal{L}_{\text{QCD}}^{\text{ext}} \right) = \int \mathcal{D}\Sigma  \exp \left(i \int d^4 x \, \mathcal{L}_{\chi\text{pt}}^{\text{ext}}\right) \, ,
\end{equation}
which can be used to construct the chiral counterparts of fermionic bilinears by taking the functional derivatives of the QCD and the $\chi$pt actions with respect to the appropriate external sources. 

For instance, the scalar Dirac bilinear $\bar{q}_i \mathcal{Z}_{ij} q_j$ can be obtained by taking a derivative of $\mathcal{L}^{\text{ext}}_{\text{QCD}}$ 
with respect to the external source $\chi$ specialized to $s = \mathcal{Z}$. Its chiral counterpart can be constructed by simply taking the same functional derivatives but acting on $\mathcal{L}^{\text{ext}}_{\chi\text{pt}}$: 
\begin{equation}
\label{eq:Scalar_Dirac_Bilinear}
\bar{q}_i \mathcal{Z}_{ij} q_j = - \mathcal{Z}_{ij}\frac{\partial \mathcal{L}^{\text{ext}}_{\text{QCD}}}{\partial\mathcal{Z}_{ij}} \equiv - \mathcal{Z}_{ij}\frac{\partial \mathcal{L}^{\text{ext}}_{\chi\text{pt}}}{\partial\mathcal{Z}_{ij}} = - \frac{f_\pi^2}{2} B_0 \text{Tr} \left[\mathcal{Z}(\Sigma + \Sigma^\dagger)\right]\,.
\end{equation}
Quantities that are obtained by applying Noether's theorem to the QCD Lagrangian have their chiral counterpart in quantities that are constructed by applying the same procedure to $\mathcal{L}_{\chi \text{pt}}$ instead.

In particular, we will need the $SU(3)_{L/R}$ currents and the energy-momentum tensor, for which we find the following expressions:
\begin{equation}
\begin{split}
j^{\mu}_{L/R} &= \frac{\partial \mathcal{L}_{\chi \text{pt}}}{\partial (\partial_\mu\Sigma)} \delta_{L/R} \Sigma +   \frac{\partial \mathcal{L}_{\chi \text{pt}}}{\partial (\partial_\mu\Sigma^\dagger)} \delta_{L/R} \Sigma^\dagger \, , \\
\theta^{\mu\nu}& = \frac{\partial \mathcal{L}_{\chi \text{pt}}}{\partial (\partial_\mu\Sigma)} \partial^\nu \Sigma +  \frac{\partial \mathcal{L}_{\chi \text{pt}}}{\partial (\partial_\mu\Sigma^\dagger)} \partial^\nu \Sigma^\dagger - \eta^{\mu\nu} \mathcal{L}_{\chi\text{pt}} \, , 
\end{split}
\end{equation}
where $\delta_L \Sigma = -i\alpha^a_L T_a \Sigma$, $\delta_R \Sigma = i \alpha^a_R \Sigma T_a$, $\delta_L \Sigma^\dagger = i \alpha^a_L \Sigma^\dagger T_a$, $\delta_R \Sigma^\dagger = - i \alpha^a_R T_a \Sigma^\dagger$, and $T_a = \lambda_a/2$ are the generators of the group $SU(3)_{L/R}$.
As a result, 
the currents of (\ref{eq:Trace_Anomaly}) 
and (\ref{eq:Completeness_Currents}) and the 
energy-momentum tensor
have the following chiral representation:
\begin{equation}
\label{eq:Chiral_Counterparts_1}
\begin{split}
& j_{V, a}^\mu = j_{R, a}^\mu + j_{L, a}^\mu = i \frac{f_\pi^2}{2} \text{Tr} \left[T_a(D^\mu \Sigma^\dagger) \Sigma + T_a(D^\mu \Sigma) \Sigma^\dagger\right] \, , \\
& j_{A, a}^\mu = j_{R, a}^\mu - j_{L, a}^\mu = i \frac{f_\pi^2}{2} \text{Tr} \left[T_a (D^\mu \Sigma^\dagger) \Sigma - T_a(D^\mu \Sigma) \Sigma^\dagger\right] \, , \\
& j_{V}^\mu = j_{R}^\mu + j_{L}^\mu =
i \frac{f_\pi^2}{2} \text{Tr} \left[(D^\mu \Sigma^\dagger) \Sigma + (D^\mu \Sigma) \Sigma^\dagger\right] \, , \\
& j_{A}^\mu = j_{R}^\mu - j_{L}^\mu = i \frac{f_\pi^2}{2} \text{Tr} \left[(D^\mu \Sigma^\dagger) \Sigma - (D^\mu \Sigma) \Sigma^\dagger\right] \, , \\
& \theta^{\mu\nu} = \frac{f_\pi^2}{4} \text{Tr} \left[\partial^\mu \Sigma^\dagger \partial^\nu \Sigma + \partial^\mu \Sigma \partial^\nu \Sigma^\dagger  \right] - \eta^{\mu\nu} \left[ \frac{f_\pi^2}{4} \text{Tr} \left(\partial^\mu \Sigma^\dagger \partial_\mu \Sigma\right) + \frac{f_\pi^2}{2} B_0 \text{Tr} \left(M_q (\Sigma + \Sigma^\dagger )\right)\right]\,.
\end{split}
\end{equation}
%

%

The bosonization procedure outlined before leads to the following chiral Lagrangian: 
\begin{equation}
\label{eq:ALP_Chipt_Lag_V1}
\begin{split}
\mathcal{L}_{\text{ALP}}^{\chi\text{pt}} 
&= \frac{\partial_\mu \phi }{\Lambda}\, 
\left[ 
2\, \text{Tr} (Y_V T_a) \, j^{\mu, a}_V +2 \, \text{Tr} (Y_A T_a)\, j^{\mu, a}_A  
+ \frac{1}{N} \text{Tr} (Y_A)\,j^{\mu}_A
\right] \\
&+ \kappa \, \frac{f_\pi^2}{2}\, \frac{\phi}{\Lambda} \,\left[ \text{Tr} (\partial^\mu \Sigma \partial_\mu \Sigma^\dagger) + 4 B_0 \text{Tr}\left[M_q (\Sigma + \Sigma^\dagger)\right] \right]
- \frac{f_\pi^2}{2} \frac{v}{\Lambda} \, B_0 \, \phi \, \text{Tr} \left[\mathcal{Z}(\Sigma + \Sigma^\dagger)\right] \\
&+e^2 \frac{c_\gamma}{\Lambda} \, \phi \,  F^{\mu\nu}F_{\mu\nu} + e^2 \frac{\tilde{c}_\gamma}{\Lambda} \, \phi \, F^{\mu\nu}\tilde{F}_{\mu\nu} 
+ \mathcal{O} \left( \frac{1}{\Lambda^2}\right) \, , 
\end{split}
\end{equation}
where $j_{V, A}^{\mu, a}$ are given in \eqref{eq:Chiral_Counterparts_1} and where we have neglected the 
ALP interactions with the diagonal vectorial quark current, being the latter a conserved current. Hereafter, we also neglect the ALP interaction with the anomalous singlet axial current, that is related to the $\eta'$ meson, as in this case the convergence of the standard $\chi \text{pt}$ approach is questionable. 
For a recent account of ALP-$\eta'$ interactions 
see however Ref.~\cite{Gao:2022xqz}.

Moreover, the Lagrangian density in \eqref{eq:ALP_Chipt_Lag_V1} has to be supplemented with the ALP-dependent mass term of the QCD Lagrangian: 

\begin{equation}
\label{eq:Mass_ALP_Chipt_Lag}
\begin{split}
\mathcal{L}_{\text{Mass}}^{\chi\text{pt}} &=  \frac{f_\pi^2}{2} B_0 \text{Tr} \left[M^\phi_q \Sigma^\dagger + \Sigma M^{\phi\dagger}_q \right] \\
&=  \frac{f_\pi^2}{2} B_0 \text{Tr} \left[M_q (\Sigma^\dagger + \Sigma)\right] + i \frac{\phi}{\Lambda} \lambda_g\frac{f_\pi^2}{2} B_0 \text{Tr} \left[\left\{M_q, Q_A\right\} (\Sigma^\dagger - \Sigma)\right]\\
&- \frac{1}{2} \frac{\phi^2}{\Lambda^2}\lambda_g^2 \frac{f_\pi^2}{2} B_0 \text{Tr} \left[\left\{\left\{M_q, Q_A\right\}, Q_A\right\} (\Sigma^\dagger + \Sigma)\right] + \mathcal{O} \left( \frac{1}{\Lambda^2}\right) \,,
\end{split}
\end{equation}
 where in the last line, among the terms with a $1/\Lambda^2$ factor, we have explicitly highlighted the leading operator that contributes to the ALP mass. 

\subsection{ALP interactions with Baryons}

In this Section, we will concisely deal with the construction of an EFT describing the interaction of baryons with mesons (see, e.g. 
\cite{Scherer:2002tk,10.1007/BFb0104294}) as well as the inclusion of ALP interactions.
The baryon octet $\mathcal{B}(x)$ is conveniently described by a $3 \times 3$ matrix transforming in the adjoint representation of $SU(3)_V$

\begin{equation}
\mathcal{B} = \begin{bmatrix} \frac{\Sigma_0}{\sqrt{2}} + \frac{\Lambda_0}{\sqrt{6}} & \Sigma^+ & p\\
\Sigma^- & -\frac{\Sigma_0}{\sqrt{2}} + \frac{\Lambda_0}{\sqrt{6}} & n\\
\Xi^- & \Xi^0 & - \frac{2}{\sqrt{6}} \Lambda_0\end{bmatrix}\,, 
\end{equation}
%
 where the baryonic multiplet transforms under $SU(3)_L\times SU(3)_R$ according to $\mathcal{B}\rightarrow K(L,R,\Sigma)\mathcal{B} K^{\dagger}(L,R,\Sigma)$, where $K(L,R,\Sigma) = \sqrt{R\Sigma L^\dagger}^{-1}R\sqrt{\Sigma}$.
Differently from the case of mesons, we can no longer consistently organize the chiral Lagrangian for relativistic baryons in a series expansion in the small parameter $p/\Lambda_{\chi}$, with $\Lambda_{\chi} = 4\pi f_\pi \approx 1$ GeV. Indeed, higher dimension operators for relativistic baryons would necessarily be of order $m_B/\Lambda_{\chi} \sim \mathcal{O}(1)$.

A consistent derivative expansion for baryons in a chiral effective theory can be nonetheless developed by treating the baryon fields as heavy fields \cite{JENKINS1991558}. The fundamental concept of this approach is that in the chiral limit the momentum transferred from one baryon to another via pion exchange is small if compared to the average baryon mass, so that the velocity of baryons can be safely considered to be conserved by such processes (it can however change as a consequence of external currents or interactions, other than strong ones). 

With this idea in mind one can then parametrize the baryon momentum 
as $p^\mu = m_B v^\mu + k^\mu$, where $v^\mu$ is the velocity of the baryon, $m_B$ its mass and $k^\mu $ the residual momentum measuring the amount by which the baryon is off-shell because of its interactions. 
In the heavy baryon limit, the Dirac propagator simplifies to
\begin{equation}
i \frac{\slashed{p} + m_B}{p^2 - m^2_B + i\epsilon} = 
i \frac{m_B\slashed{v} + m_B + \slashed{k}}{2m_B v\cdot k + k^2 + i\epsilon} 
\rightarrow ~ 
i\frac{1 + \slashed{v}}{2 v\cdot k + i\epsilon} \, , 
\end{equation}
where in the last step we have expanded for $k \ll m_B$.
Hence, the propagator of the heavy baryon contains a velocity-dependent projection operator, $(1+\slashed{v})/2$, and it is convenient to decompose the original quark field $\mathcal{B}(x)$
as follows
\begin{equation}
    \mathcal{B}(x) = e^{-i m_B v \cdot x}[B_v(x) + H_v(x)] \, , 
    \label{eq:HQEFT_expansion}
\end{equation}
where 
\begin{equation}
B_v(x) = e^{i m_B v \cdot x}\frac{1+ \slashed{v}}{2}\mathcal{B}(x)\,, \qquad
H_v(x) = e^{i m_B v \cdot x}\frac{1- \slashed{v}}{2}\mathcal{B}(x) \, , 
\end{equation}
and $B_v(x)$ can be identified as the definite-velocity (light-component) baryon field. The exponential prefactor subtracts $m_B v^\mu$ from the heavy quark momentum. 
As a result, the $B_v(x)$ field induces effects at leading order while the effects of $H_v(x)$ are suppressed by powers of $1/m_B$. Finally, neglecting $H_v(x)$  and substituting Eq.~(\ref{eq:HQEFT_expansion}) 
into the Lagrangian for the heavy baryon field, $\bar{\mathcal{B}} (i\slashed{D} - m_B)\mathcal{B}$, 
gives $\bar{B}_v i\slashed{D}B_v$.


The interactions of baryons with mesons are accounted for by introducing the quantities
\begin{equation}
\label{eq:Miscellanea_Baryon_Lagrangian}
\begin{split}
\xi (x) &= \exp \left[ i \frac{\pi(x)}{2 f_\pi}\right] \, ,\quad \text{trasforming as} \ \xi \rightarrow \sqrt{R\Sigma L^\dagger} = R \xi K^\dagger(L,R,\Sigma) \, , \\
\mathcal{A}^\mu &= \frac{i}{2} (\xi \partial^\mu \xi^\dagger - \xi^\dagger \partial^\mu \xi) = \frac{\partial^\mu \pi}{2 f_\pi} + \dots \, , \\
\mathcal{V}^\mu &= \frac{1}{2} (\xi \partial^\mu \xi^\dagger + \xi^\dagger \partial^\mu \xi) = \frac{1}{8} \frac{\left[\pi, \partial^\mu \pi\right]}{f_\pi^2} + \dots \, , 
\end{split}
\end{equation}
which allow one to build the leading-order heavy baryon Lagrangian \cite{Donoghue:1992dd,Ecker:1994gg}: 
\begin{equation}
\label{eq:Heavy_Baryon_Lagrangian}
    \mathcal{L}_{\text{HB}} = i \text{Tr} (\bar{B}_v \gamma^\mu D_\mu B_v) - D \, \text{Tr} (\bar{B}_v \gamma^\mu \gamma_5 \left\{\mathcal{A}_\mu, B_v \right\}) -  F \, \text{Tr} (\bar{B}_v \gamma^\mu \gamma_5 \left[\mathcal{A}_\mu, B_v \right]) \, ,
\end{equation}
where the covariant derivative is defined as 
$D_\mu X = \partial_\mu X + \left[\mathcal{V}_\mu, X\right]$.  
Noether's currents for baryons can be constructed by following 
the usual procedure, yielding: 
\begin{equation}
\label{eq:Baryonic_Currents}
\begin{split}
(j^{a, \mu}_{A})_{\text{B}} &= \frac{1}{2} \text{Tr} \left(\bar{B}_v \gamma^\mu \left[\xi T^a \xi^\dagger - \xi^\dagger T^a \xi, B_v\right]\right)+ \frac{D}{2} \text{Tr} \left(\bar{B}_v \gamma^\mu \gamma_5 \left\{\xi T^a \xi^\dagger + \xi^\dagger T^a \xi, B_v\right\}\right)\\
& \qquad + \frac{F}{2} \text{Tr} \left(\bar{B}_v \gamma^\mu \gamma_5\left[\xi T^a \xi^\dagger + \xi^\dagger T^a \xi, B_v\right]\right)\,,\\
(j^{a, \mu}_{V})_{\text{B}} &= -\frac{1}{2} \text{Tr} \left(\bar{B}_v \gamma^\mu \left[\xi T^a \xi^\dagger + \xi^\dagger T^a \xi, B_v\right]\right)- \frac{D}{2} \text{Tr} \left(\bar{B}_v \gamma^\mu \gamma_5 \left\{\xi T^a \xi^\dagger - \xi^\dagger T^a \xi, B_v\right\}\right)\\
& \qquad - \frac{F}{2} \text{Tr} \left(\bar{B}_v \gamma^\mu \gamma_5 \left[\xi T^a \xi^\dagger - \xi^\dagger T^a \xi, B_v\right]\right)\,,
\\
(j^{\mu}_{A})_{\text{B}} &= \frac{1}{2} \text{Tr} \left(\bar{B}_v \gamma^\mu \left[\xi \xi^\dagger - \xi^\dagger \xi, B_v\right]\right)+ \frac{D}{2} \text{Tr} \left(\bar{B}_v \gamma^\mu \gamma_5 \left\{\xi\xi^\dagger + \xi^\dagger \xi, B_v\right\}\right)\\
& \qquad + \frac{F}{2} \text{Tr} \left(\bar{B}_v \gamma^\mu \gamma_5\left[\xi \xi^\dagger + \xi^\dagger \xi, B_v\right]\right)\,,\\
(j^{\mu}_{V})_{\text{B}} &= -\frac{1}{2} \text{Tr} \left(\bar{B}_v \gamma^\mu \left[\xi \xi^\dagger + \xi^\dagger \xi, B_v\right]\right)- \frac{D}{2} \text{Tr} \left(\bar{B}_v \gamma^\mu \gamma_5 \left\{\xi \xi^\dagger - \xi^\dagger \xi, B_v\right\}\right)\\
& \qquad - \frac{F}{2} \text{Tr} \left(\bar{B}_v \gamma^\mu \gamma_5 \left[\xi \xi^\dagger - \xi^\dagger \xi, B_v\right]\right)\,.
\end{split}
\end{equation}
In a two-flavour setting the fundamental baryonic fields are the proton and the neutron, combined in the nucleon doublet $\mathcal{N}  =(p, n)^T$  transforming under $SU(2)_L\times SU(2)_R$ as $\mathcal{N} \rightarrow K(L,R,\Sigma) \mathcal{N}$. Its heavy-baryon counterpart is defined as before, $N_v (x) = \exp(im_N v \cdot x)\frac{1+ \slashed{v}}{2}\mathcal{N}$ and the leading-order Lagrangian is found to be \cite{Donoghue:1992dd, Ecker:1994gg}

\begin{equation}
\label{eq:Nucleon_Lagrangian}
    \mathcal{L}_{\text{HN}} = i \bar{N}_v \gamma^\mu D_\mu N_v - g_A \bar{N}_v \gamma^\mu \gamma_5 \mathcal{A}_\mu N_v \, , 
\end{equation}
where now $D_\mu = \partial_\mu + \mathcal{V}_\mu$.
In this case the Noether's currents read: 
\begin{equation}
\label{eq:Nucleonic_Currents}
\begin{split}
(j^{a, \mu}_{A})_{\text{N}} &= \frac{1}{2} \bar{N}_v \gamma^\mu \left[\xi T^a \xi^\dagger - \xi^\dagger T^a \xi\right] N_v + \frac{g_A}{2} \bar{N}_v \gamma^\mu \gamma_5\left[\xi T^a \xi^\dagger + \xi^\dagger T^a \xi \right] N_v\,,\\
(j^{a, \mu}_{V})_{\text{N}} &= -\frac{1}{2} \bar{N}_v \gamma^\mu \left[\xi T^a \xi^\dagger + \xi^\dagger T^a \xi\right] N_v - \frac{g_A}{2} \bar{N}_v \gamma^\mu \gamma_5 \left[\xi T^a \xi^\dagger - \xi^\dagger T^a \xi \right] N_v\,,
\\
(j^{\mu}_{A})_{\text{N}} &= \frac{1}{2} \bar{N}_v \gamma^\mu \left[\xi \xi^\dagger - \xi^\dagger \xi\right] N_v + \frac{g_A}{2} \bar{N}_v \gamma^\mu \gamma_5\left[\xi \xi^\dagger + \xi^\dagger \xi \right] N_v \,,\\
(j^{\mu}_{V})_{\text{N}} &= -\frac{1}{2} \bar{N}_v \gamma^\mu \left[\xi \xi^\dagger + \xi^\dagger \xi\right] N_v - \frac{g_A}{2} \bar{N}_v \gamma^\mu \gamma_5 \left[\xi \xi^\dagger - \xi^\dagger \xi \right] N_v
\,.
\end{split}
\end{equation}
 Note that the singlet currents need to be supplemented by an extra term arising within an 
$U(3)_L \times U(3)_R$ chiral formalism. 
For our purposes, this contribution 
results into a modified coupling of the singlet axial currents, which can be parametrized at leading order as 
\begin{align}
\label{eq:jANp}
(j^\mu_A)'_{\text{N}} &= g_0 \, \bar{N}_v \gamma^\mu \gamma_5 N_v \, , \\
\label{eq:jABp}
(j^\mu_A)'_{\text{B}} &= S \, \text{Tr} ( \bar{B}_v \gamma^\mu \gamma_5 B_v ) \, , 
\end{align}
respectively in the $N_f = 2$ and 3 setting. 


The ALP interactions with baryons can be obtained starting from \eqref{eq:AAA} and exploiting the hadron-quark duality as well as the quark rotation \eqref{eq:Quark_Field_Redefinition} to get rid of the term $\phi G \tilde{G}$. As a result, we obtain 
\begin{equation}
\label{eq:QCD_Scale_ALP_Lag_V2}
\begin{split}
\mathcal{L}_{\text{ALP}}^{\text{QCD}} &= \frac{\partial_\mu \phi }{\Lambda}\, 
\left[  
2\, \text{Tr} (Y_V T_a) \, (j_{V, a}^\mu)_{\text{B}} + 
\frac{1}{N}\, \text{Tr} (Y_V) \, (j_{V}^\mu)'_{\text{B}} + 
2\, \text{Tr} (Y_A T_a) \, (j_{A, a}^\mu)_{\text{B}} + 
\frac{1}{N}\, \text{Tr} (Y_A) \, (j_{A}^\mu)'_{\text{B}}
\right] \\
& + \frac{v}{\Lambda} \, \phi \, \bar{q} y_S q 
+ g^2_s \frac{C_g}{\Lambda}\,\phi\, G^{a}_{\mu\nu}G_{a}^{\mu\nu}
+e^2 \frac{c_\gamma}{\Lambda} \, \phi \,  F^{\mu\nu}F_{\mu\nu} + e^2 \frac{\tilde{c}_\gamma}{\Lambda} \, \phi \, F^{\mu\nu}\tilde{F}_{\mu\nu}\,.
\end{split}
\end{equation}
Note that the trace anomaly condition \eqref{eq:Trace_Anomaly} has not been 
employed so far to remove the term $\phi GG$.
The hadronization of such term in the framework 
of heavy baryons will be discussed in the next Sections.
Moreover, in contrast to the case of the chiral Lagrangian for mesons, there are now no interaction terms stemming 
from heavy baryons masses as they are suppressed by
powers of $1/m_B$.

\section{ALP interactions for $N_f =2$}
\label{sec:ALPINT2F}

The general results presented in the previous Section can now be specialized to the low-energy setting of interest. 
Hereafter, we will focus on a two-flavours setting ($N_f=2$) 
where the strange quark and its phenomenological impact can be safely neglected.
Throughout this Section, we take $Q_V = \text{diag}\,(q^u_V,q^d_V)$ and $Q_A = \text{diag}\,(q^u_A,q^d_A)$.

\subsection{ALP interactions with Mesons}


We are now ready to expand Eq.~\eqref{eq:ALP_Chipt_Lag_V1} and \eqref{eq:Mass_ALP_Chipt_Lag} in terms of the constituent meson fields. Keeping ALP interactions with no more than 
three mesons, and employing the Gell-Mann Okubo relation 
$B_0 = m_\pi^2/(m_u + m_d)$, we find
\begin{equation}
\label{eq:ALP_Chipt_Lag_2f_V1}
\begin{split}
\mathcal{L}_{\text{ALP}}^{\chi\text{pt}} &= \kappa \, \frac{\phi}{\Lambda} \, \left[\partial^\mu \pi_0 \partial_\mu \pi_0 + 2 \partial^\mu \pi^+ \partial_\mu \pi^- - 2 m_\pi^2 (\pi_0^2 + 2 \pi^+ \pi^-)\right] + \frac{v}{2} \frac{\phi}{\Lambda} m_\pi^2 \frac{\mathcal{Z}_u + \mathcal{Z}_d}{m_u + m_d} (\pi_0^2 + 2 \pi^+ \pi^-) \\
&+ (Y_A^u - Y_A^d)\frac{f_\pi}{\Lambda}\,  \partial^\mu \phi\,  \partial_\mu \pi_0  + 2 \frac{(Y_A^u - Y_A^d)}{3 f_\pi \Lambda} \, \partial^\mu \phi\,   (\pi_0 \pi^- \partial_\mu \pi^+ + \pi_0 \pi^+ \partial_\mu \pi^- - 2\pi^+ \pi^- \partial_\mu \pi^0) \\
&+ i \frac{(q_V^u - q_V^d)}{\Lambda}\,  \partial^\mu \phi\,  (\pi^-D_\mu \pi^+  - \pi^+D_\mu \pi^-)\,,
\end{split}
\end{equation}
where $D_\mu \pi^\pm = (\partial_\mu \pm i e A_\mu)\pi^\pm$.
Moreover, from the mass term we obtain the following interactions
\begin{equation}
\label{eq:Mass_ALP_Lag_2f_V1}
\mathcal{L}_{\text{m}}^{\chi\text{pt}} =  
2 m_\pi^2 \lambda_g \,\frac{m_u q_u - m_d q_d}{m_u + m_d}
\, \frac{f_\pi}{\Lambda}
\left[
\phi \, \pi_0 - \frac{1}{6 f^2_\pi}\, \phi \, \pi_0 (\pi^2_0 + 2 \pi^+ \pi^-) 
\right]
- 2m_\pi^2\lambda_g^2\frac{f_\pi^2}{\Lambda^2} \frac{m_uq_u^2 + m_d q_d^2}{m_u + m_d}\, \phi^2\,.
\end{equation}
Non-perturbative effects generate 
a mass term for the ALP even if it was originally absent, 
in analogy to the case of the QCD axion.
%
%
Eqs.~\eqref{eq:ALP_Chipt_Lag_2f_V1} and \eqref{eq:Mass_ALP_Lag_2f_V1} contain both a kinetic mixing and a mass mixing term coupling the ALP to the neutral pion. These can be schematically written as 

\begin{equation}
 \mathcal{L}^{\text{ALP mixing}}_{\chi \text{pt}} = \frac{1}{2} \partial^\mu\varphi^T \, \mathbf{Z} \, \partial_\mu \varphi - \frac{1}{2} \varphi^T \, \mathbf{M}\, \varphi \, ,  \qquad \text{with} \qquad \varphi = \begin{pmatrix}\phi \\ \pi_0\end{pmatrix} \, ,
\end{equation}
where 
\begin{equation}
\begin{split}
&\mathbf{Z} = \begin{bmatrix} 1 & \epsilon \\ \epsilon & 1\end{bmatrix}\,, 
\qquad \qquad\qquad
\mathbf{M} = \begin{bmatrix} m_\phi^2 & -\epsilon\, \alpha \\ -\epsilon \, \alpha & m_\pi^2\end{bmatrix} \, , 
\end{split} 
\end{equation}
and where we have defined
\begin{equation}
\epsilon = (Y_A^u - Y_A^d)\frac{f_\pi}{\Lambda}\,, \quad
\alpha = 2 \frac{m_\pi^2}{(Y_A^u - Y_A^d)} \lambda_g\frac{m_u q^u_A - m_d q_A^d}{m_u + m_d}\,,\quad m_\phi^2  = M_\phi^2 + 4 m_\pi^2 (\lambda_g)^2 \frac{f_\pi^2}{\Lambda^2} \frac{m_u (q_A^{u})^2 + m_d (q_A^{d})^2}{m_u + m_d}\,.
\end{equation}
Here, we have included a potential explicit mass term for the ALP, denoted as $M_\phi$.
After shifting the field doublet $\varphi$ to make the kinetic terms canonical and diagonalising the mass matrix ${\mathbf{M}}$, we end up with the following 
Lagrangian
\begin{equation}
\label{eq:Scheme_Dependent_Lag_2f}
\begin{split}
\mathcal{L}_{\text{ALP+ Mass}}^{\chi\text{pt}} &=  - \frac{m^2_{\phi}}{2} \phi^2 + \kappa \, \frac{\phi}{\Lambda} \, \left[\partial^\mu \pi_0 \partial_\mu \pi_0 + 2 \partial^\mu \pi^+ \partial_\mu \pi^- - 2 m_\pi^2 (\pi_0^2 + 2 \pi^+ \pi^-)\right]\\
&+ \frac{v}{2} \frac{\phi}{\Lambda} m_\pi^2 \frac{\mathcal{Z}_u + \mathcal{Z}_d}{m_u + m_d} (\pi_0^2 + 2 \pi^+ \pi^-) -  \alpha \frac{Y_A^u-Y_A^d}{6 f_\pi \Lambda}\,  \phi \, \pi_0 (\pi_0^2 + 2 \pi^+ \pi^-)\\
&+ 2 \frac{(Y_A^u - Y_A^d)}{3 f_\pi \Lambda} \, \partial^\mu \phi\,   (\pi_0 \pi^- \partial_\mu \pi^+ + \pi_0 \pi^+ \partial_\mu \pi^- - 2\pi^+ \pi^- \partial_\mu \pi^0) \\
&+ i \frac{(q_V^u - q_V^d)}{\Lambda}\,  \partial^\mu \phi\,  (\pi^-D_\mu \pi^+  - \pi^+D_\mu \pi^- ) +e^2 \frac{c_\gamma}{\Lambda} \, \phi \,  F^{\mu\nu}F_{\mu\nu} + e^2 \frac{\tilde{c}_\gamma}{\Lambda} \, \phi \, F^{\mu\nu}\tilde{F}_{\mu\nu} \, , 
\end{split}
\end{equation}
where in the above expression the fields and masses are the physical ones, defined as\footnote{Note that this formula is valid for $|m_\pi^2-m_\phi^2| \gg m_\pi^2 f_\pi / \Lambda$, that is satisfied barring a fine-tuned region of parameter space.} 
\begin{equation}
\label{eq:Physical_Fields_2f}
\begin{split}
    &\phi \rightarrow \phi + \epsilon \frac{m_\pi^2 + \alpha}{m_\pi^2-m_\phi^2} \pi_0\,,
    \qquad\quad~ 
    m_\phi^2 \rightarrow m_\phi^2 - \epsilon^2 \frac{(m_\phi^2 + \alpha)^2}{m_\pi^2 - m_\phi^2}\,,
    \\
    &\pi_0 \rightarrow \pi_0 - \epsilon \frac{m_\phi^2 + \alpha}{m_\pi^2-m_\phi^2} \phi\,,
    \qquad\quad 
    m_\pi^2 \rightarrow m_\pi^2 + \epsilon^2 \frac{(m_\pi^2 + \alpha)^2}{m_\pi^2 - m_\phi^2}\,.
\end{split}
\end{equation}
Note that the kinetic and mass mixings between the neutral pion and the ALP induce also a modification of the following four-pion interactions
\begin{align}
\mathcal{L}_{4\pi} &= \frac{m_\pi^2}{24 f_\pi^2} (\pi_0^4 + 4 \pi_0^2 \pi^+ \pi^- + 4 \pi^+ \pi^+ \pi^-\pi^-) + \frac{1}{6 f_\pi^2} \left[ \pi^- \pi^-D^\mu \pi^+ D_\mu \pi^+ + \pi^+ \pi^+D^\mu \pi^- D_\mu \pi^- 
- 2 \pi^+ \pi^- \partial^\mu \pi_0 \partial_\mu \pi^0
\right.\nonumber\\
&\left. + 2 \pi_0 \partial^\mu \pi_0 (\pi^-D_\mu \pi^+ + \pi^+D_\mu \pi^-) - 2 D^\mu \pi^+D_\mu \pi^- (\pi_0^2 + \pi^+ \pi^-)\right] + \frac{e^2}{16\pi^2}\frac{\pi_0}{f_\pi} F^{\mu\nu}\tilde{F}_{\mu\nu} \, , 
\end{align}
which leads to the additional ALP interaction terms
\begin{equation}
\begin{split}
\mathcal{L}^{\text{ALP}}_{4\pi} &= - \epsilon \frac{m_\phi^2 + \alpha}{m_\pi^2-m_\phi^2} \left\{\frac{e^2}{16\pi^2}\frac{\phi}{f_\pi} F^{\mu\nu}\tilde{F}_{\mu\nu}+\frac{m_\pi^2}{6 f_\pi^2} ( \phi \,\pi_0^3 + 2 \phi\,  \pi_0 \pi^+ \pi^-) - \frac{1}{3 f_\pi^2} \left[  - 2 \pi^+ \pi^- \partial^\mu \phi \partial_\mu \pi^0 \right.\right.\\
&\left. \left.+ \, \phi \, \partial^\mu \pi_0 (\pi^-D_\mu \pi^+ + \pi^+D_\mu \pi^-)+  \pi_0 \partial^\mu \phi (\pi^-D_\mu \pi^+ + \pi^+D_\mu \pi^-) - 2 \phi \, \pi_0\, D^\mu \pi^+D_\mu \pi^- \right] \right\}\,.
\end{split}
\end{equation}
A common choice in the literature for the parameters 
$q_A^u$ and $q_A^d$ is \cite{Georgi:1986df}
\begin{equation}
 q_A^u = \frac{1}{2}\frac{m_d}{m_u + m_d} \qquad \text{and} \qquad  
 q_A^d = \frac{1}{2}\frac{m_u}{m_u + m_d} \, , 
\end{equation}
which corresponds to $\alpha = 0$. However, as noted in \cite{Bauer:2020jbp}, a more convenient 
choice corresponds to set $\alpha = -m_\phi^2$ as in this case the interactions in $\mathcal{L}^{\text{ALP}}_{4\pi}$ vanish. 
This choice can be implemented by setting
\begin{equation}
 q_A^u = \frac{1}{2}\frac{m_d}{m_u + m_d} - \frac{m_\phi^2}{m_\pi^2-m_\phi^2} \frac{\Delta^A_{ud}}{2 \lambda_g}\qquad \text{and} \qquad  
 q_A^d = \frac{1}{2}\frac{m_u}{m_u + m_d} + \frac{m_\phi^2}{m_\pi^2-m_\phi^2} \frac{\Delta^A_{ud}}{2 \lambda_g} \, , 
\end{equation}
where we have defined the coupling constant
\begin{equation}
\Delta^A_{ud} =  Y_P^u  - Y_P^d  -  \frac{\lambda_g}{2} \frac{m_d-m_u}{m_d+m_u} = \frac{m_\pi^2-m_\phi^2}{m_\pi^2} (Y_A^u-Y_A^d)\,.
\end{equation}
In this case the ALP-Meson interaction Lagrangian simply reduces to 
\begin{equation}
\label{eq:Simplest_Lag_2f}
\begin{split}
\mathcal{L}_{\text{ALP+ Mass}}^{\chi\text{pt}} &=  - \frac{m^2_\phi}{2} \phi^2 + \kappa \, \frac{\phi}{\Lambda} \, \left[\partial^\mu \pi_0 \partial_\mu \pi_0 + 2 \partial^\mu \pi^+ \partial_\mu \pi^-\right]\\
&+ \frac{\phi}{\Lambda}\,  (\pi_0^2 + 2\pi^+ \pi^-) m_\pi^2 \left[\frac{v}{2} \frac{\mathcal{Z}_u + \mathcal{Z}_d}{m_u + m_d} - 2 \, \kappa+ \pi_0\frac{m_\phi^2}{m_\pi^2-m_\phi^2}\frac{\Delta^A_{ud}}{6 f_\pi} \right]  \\
&+ \frac{2}{3}\frac{\Delta^A_{ud}}{f_\pi \Lambda} \frac{m_\pi^2}{m_\pi^2- m_\phi^2} \, \partial^\mu \phi\,   (\pi_0 \pi^- \partial_\mu \pi^+ + \pi_0 \pi^+ \partial_\mu \pi^- - 2\pi^+ \pi^- \partial_\mu \pi^0) \\
&+ i \frac{(q_V^u - q_V^d)}{\Lambda}\,  \partial^\mu \phi\,  (\pi^-D_\mu \pi^+  - \pi^+D_\mu \pi^- ) +e^2 \frac{c_\gamma}{\Lambda} \, \phi \,  F^{\mu\nu}F_{\mu\nu} + e^2 \frac{\tilde{c}_\gamma}{\Lambda} \, \phi \, F^{\mu\nu}\tilde{F}_{\mu\nu} \, , 
\end{split} 
\end{equation}
where now
\begin{equation}
\begin{split}
m^2_{\phi} &= M_\phi^2 + m_\pi^2 (\lambda_g)^2\frac{m_um_d}{(m_u + m_d)^2} \frac{f_\pi^2}{\Lambda^2} + m_\pi^2 \frac{M_\phi^4 }{(m_\pi^2-M_\phi^2)^2}(\Delta^A_{ud})^2\frac{f_\pi^2}{\Lambda^2} + \mathcal{O} \left(\frac{1}{\Lambda^4}\right)\,.
\end{split}
\end{equation}

\subsection{ALP interactions with Baryons}

The ALP interactions with nucleons stem from the derivative couplings to vectorial and axial currents, the direct couplings to the QCD energy-momentum tensor, and the quark bilinear $\bar{q}\mathcal{Z}q$, 
see Eq.~\eqref{eq:QCD_Scale_ALP_Lag_V2}. 

Focusing on the derivative interactions of an ALP with two nucleons, 
we find the following Lagrangian at leading order:
\begin{equation}
\label{eq:Derivative_Couplings_To_Nucleons}
\begin{split}
\mathcal{L}_{\text{ALP}}^{\partial N} &= \frac{\partial_\mu \phi}{\Lambda} \left[ \text{Tr}(Y_A \sigma_3) (j_{A,3}^{\mu})_{\text{N}} +\frac{1}{2} \text{Tr}(Y_A) (j_{A}^\mu)'_{\text{N}} \right] \\
& =\frac{\partial_\mu \phi}{\Lambda} \left[ \frac{g_A}{2}  (Y_A^u-Y_A^d)(\bar{p}_v \gamma^\mu \gamma_5 p_v -\bar{n}_v \gamma^\mu \gamma_5 n_v)  + \frac{g_0}{2} (Y_A^u+Y_A^d) (\bar{p}_v \gamma^\mu \gamma_5 p_v +\bar{n}_v \gamma^\mu \gamma_5 n_v)\right] \,.
\end{split}
\end{equation}
In order to obtain the couplings between one ALP, two nucleons and any desired number of pions it is then enough to expand up to next-to-leading order the mesonic terms $\xi$ and $\xi^\dagger$ appearing in the definitions in Eq.~\eqref{eq:Nucleonic_Currents}, 
and include as well the expression of the singlet current in Eq.~(\ref{eq:jANp}).

Now, the coefficients $g_0$ and $g_A$ appearing in Eq.~\eqref{eq:Derivative_Couplings_To_Nucleons} can be rewritten in terms of the spin component of a given quark type in the nucleon, defined as $s^\mu \Delta_q = \bra{N} \bar{q} \gamma^\mu \gamma_5 q \ket{N}$, where $N = \{p, n\}$.
In order to do so, we recall that $2s^\mu = \bra{N} \bar{N}_v \gamma^\mu \gamma_5 N_v \ket{N}$ and match the single-nucleon matrix elements in the quark-axion Lagrangian \eqref{eq:QCD_Scale_ALP_Lag_V2} and in the effective ALP-nucleon Lagrangian \eqref{eq:Derivative_Couplings_To_Nucleons}, 
that is $\bra{N} \mathcal{L}_{\text{ALP}}^{\text{Nuc., der.}}\ket{N} = \bra{N}\mathcal{L}_{\text{ALP}}^{\text{QCD}}\ket{N}$ \cite{GrillidiCortona:2015jxo, DiLuzio:2020wdo}. 

The outcome of this procedure sets $g_0 = \Delta_u + \Delta_d$ and $g_A = \Delta_u-\Delta_d$, allowing us to write down the derivative coupling of ALPs with nucleons in a compact, matrix form:
\begin{equation}
\label{eq:LALPDN}
\mathcal{L}_{\text{ALP}}^{\partial N} = \frac{\partial_\mu \phi}{\Lambda} \bar{N}_v \tilde{C}_{\phi\text{N}}\gamma^\mu \gamma_5 N_v\,.
\end{equation}
Here, defining $\tilde{C}_{\phi\text{N}} = \text{diag} (\tilde{C}_{\phi\text{p}}, \tilde{C}_{\phi\text{n}})$ we find 
\begin{equation}
\begin{split}
\tilde{C}_{\phi\text{p}} &= Y_A^u \Delta_u + Y_A^d \Delta_d = Y_P^u \Delta_u + Y_P^d \Delta_d - \frac{\lambda_g}{2} \left[ \frac{m_u \Delta_d}{m_u + m_d} + \frac{m_d \Delta_u}{m_u + m_d} - \frac{m_\phi^2}{m_\pi^2-m_\phi^2} \frac{\Delta_u-\Delta_d}{\lambda_g} \Delta^A_{ud}\right] 
\, , \\
\tilde{C}_{\phi\text{n}} &= Y_A^u \Delta_d + Y_A^d \Delta_u = Y_P^u \Delta_d + Y_P^d \Delta_u - \frac{\lambda_g}{2} \left[ \frac{m_u \Delta_u}{m_u + m_d} + \frac{m_d \Delta_d}{m_u + m_d} + \frac{m_\phi^2}{m_\pi^2-m_\phi^2} \frac{\Delta_u-\Delta_d}{\lambda_g} \Delta^A_{ud}\right] 
\, , 
\end{split}
\end{equation}
where 
$\Delta_u = 0.858(22)$ and 
$\Delta_d = -0.418(22)$  
\cite{DiLuzio:2023tqe}. 

There is also a non-derivative type of interaction of the ALP with nucleons, given by
\begin{equation}
\label{eq:ALP_Nucleon_Interactions}
\mathcal{L}_{\text{ALP}}^N = \frac{\phi}{\Lambda} C_{\phi \text{NN}} \bar{N}_v N_v \, , 
\end{equation}
which originates from CP-even couplings of the ALP field with either quarks or gluons \cite{Cheng:1988im, Cheng:2012qr}. 
Since the ALP has no impact on the hadronic matrix element, we can then simply determine the effective coupling between an ALP and nucleons by requiring the equivalence 
\begin{equation}
    \sum_q \zeta_q \bar{q} q \,\phi +  g_{\phi gg}\, \phi \, GG  \qquad \longrightarrow \qquad   \bra{N}\sum_q \zeta_q \bar{q} q + g_{\phi gg} \, GG \ket{N}\,   \phi \,  \bar{N} N \, , 
\end{equation}
and computing the hadronic matrix element of the scalar quark and gluon densities.
This can be achieved by applying the relations
\begin{equation}
\begin{split}
&\bra{p} \bar{u} u \ket{p} = \frac{\sigma_u}{m_u} \, , \qquad \bra{p} \bar{d} d \ket{p} = \frac{\sigma_d}{m_d} \, , \qquad \bra{p} \bar{s} s \ket{p} = \frac{\sigma_s}{m_s} \, , \\
& \qquad \qquad \bra{p} GG \ket{p} = -\frac{8\pi}{9\alpha_s} (m_p- \sigma_u- \sigma_d - \sigma_s) \, , 
\end{split}
\end{equation}
to the Lagrangian in Eq.~\eqref{eq:AAA}, obtaining the effective scalar coupling of an ALP to protons:
\begin{equation}
C_{\phi \text{pp}} =  \frac{v}{m_u} y_{q,S}^u \, \sigma_u +  \frac{v}{m_d} y_{q,S}^d \, \sigma_d +  \frac{v}{m_s} y_{q,S}^s\,  \sigma_s - \frac{32\pi^2 C_g}{9} \,  (m_p - \sigma_u -\sigma_d-\sigma_s) \,.
\end{equation}
Due to iso-spin symmetry the difference between $C_{\phi \text{pp}}$ and $C_{\phi \text{nn}}$ is negligible and the scalar coupling to nucleons can be approximated by a universal value $C_{\phi \text{NN}} \simeq C_{\phi \text{pp}}$. 
Explicit values for the various quantities as well as the procedure employed in order to obtain them can be found in \cite{Cheng:2012qr} and references therein.

Clearly, due to the ALP-neutral pion mixing, one should take into account also the ALP-nucleon interactions that originate from the pion-nucleon ones upon rotation to the physical basis. More explicitly, from Eq.~\eqref{eq:Nucleon_Lagrangian} one has the following pion-nucleon interaction: 
\begin{equation}
\mathcal{L}_{N\pi} \supset -g_A \bar{N}_v \gamma^\mu \gamma_5 \frac{\partial_\mu \pi}{2 f_\pi} N_v \supset -\frac{g_A}{2f_\pi} \partial_\mu \pi_0 (\bar{p}_v \gamma^\mu \gamma_5 p_v - \bar{n}_v \gamma^\mu \gamma_5 n_v) \, , 
\end{equation}
where we stopped at order $\mathcal{O}(\pi \bar{N}N)$ in the expansion in Eq.~\eqref{eq:Miscellanea_Baryon_Lagrangian}. 

The mixing described in Eq.~\eqref{eq:Physical_Fields_2f} then necessarily shifts the couplings of the ALP to nucleons according to
\begin{equation}
\delta \mathcal{L}_{N\pi} =+\frac{g_A}{2f_\pi} \varepsilon \frac{m_\phi^2 + \alpha}{m_\pi^2-m_\phi^2} \, \partial_\mu \phi \, (\bar{p}_v \gamma^\mu \gamma_5 p_v - \bar{n}_v \gamma^\mu \gamma_5 n_v) \, .
\end{equation}
Once again, however, the choice $\alpha = -m_\phi^2$ allows us to get rid of this contribution.

\subsection{Summary and Jarlskog invariants}
\label{subsection:summary_Nf=2}

Throughout this Section we have constructed the low-energy effective Lagrangian describing the interaction of an ALP with light particles: photons, pions and nucleons. We have found a Lagrangian that can be decomposed into two pieces featuring opposite CP transformation properties, 
namely
\begin{equation}
\label{eq:Simplest_Lag_2f_CPE}
\begin{split}
\left(\mathcal{L}_{\text{ALP}}^{\chi\text{pt}}\right)^{\rm CP-even} &= e^2 \frac{c_\gamma}{\Lambda} \, \phi \,  F^{\mu\nu}F_{\mu\nu} + \frac{v}{\Lambda} y_{\ell, S}^{ij}\,  \phi \, \bar{\ell}_i \ell_j +  \kappa \, \frac{\phi}{\Lambda} \, \left[\partial^\mu \pi_0 \partial_\mu \pi_0 + 2 \partial^\mu \pi^+ \partial_\mu \pi^-\right]\\
&+ \frac{\phi}{\Lambda}\,  (\pi_0^2 + 2\pi^+ \pi^-) m_\pi^2 \left[\frac{v}{2} \frac{\mathcal{Z}_u + \mathcal{Z}_d}{m_u + m_d} - 2 \, \kappa \right]  + \frac{\phi}{\Lambda} C_{\phi \text{NN}} \bar{N}_v N_v
\end{split}
\end{equation}
and
\begin{equation}
\label{eq:Simplest_Lag_2f_CPO}
\begin{split}
\left(\mathcal{L}_{\text{ALP}}^{\chi\text{pt}}\right)^{\rm CP-odd} &=  e^2 \frac{\tilde{c}_\gamma}{\Lambda} \, \phi \, F^{\mu\nu}\tilde{F}_{\mu\nu}+ i \frac{v}{\Lambda} y_{\ell, P}^{ij}\,  \phi \, \bar{\ell}_i \gamma_5 \ell_j+ \frac{\phi}{\Lambda}\, \pi_0 (\pi_0^2 + 2\pi^+ \pi^-) m_\pi^2 \left[ \frac{m_\phi^2}{m_\pi^2-m_\phi^2}\frac{\Delta^A_{ud}}{6 f_\pi} \right] \\
&+ \frac{2}{3}\frac{\Delta^A_{ud}}{f_\pi \Lambda} \frac{m_\pi^2}{m_\pi^2- m_\phi^2} \, \partial^\mu \phi\,   (\pi_0 \pi^- \partial_\mu \pi^+ + \pi_0 \pi^+ \partial_\mu \pi^- - 2\pi^+ \pi^- \partial_\mu \pi^0)  + \frac{\partial_\mu \phi}{\Lambda} \bar{N}_v \tilde{C}_{\phi\text{N}}\gamma^\mu \gamma_5 N_v \, , 
\end{split} 
\end{equation}
where explicit expressions for the parameters appearing in the above Lagrangians can be found throughout the previous Sections. 
Note that, regardless of the scalar or pseudoscalar nature of the ALP field, CP is necessarily violated if at 
least one CP-even and CP-odd coupling is simultaneously present.

A particularly convenient way to account for CP-violating effects is to employ Jarlskog invariants. See Ref.~\cite{Bonnefoy:2022rik} for a detailed analysis at the quark level.   
In our case, they are reported in Table \ref{tab:2F_Jarlskog}.
To simplify their interpretation, we also report below the compositions of all the coefficients 
appearing in Table \ref{tab:2F_Jarlskog} in terms of the 
microscopic parameters of Eq.~\ref{eq:AAA}
%
\begin{align}
    c_\gamma  =& \, C_\gamma - \frac{\beta^0_{\text{QED}}}{\beta^0_{\text{QCD}}}  C_g \, ,\\
    \tilde{c}_\gamma =& \, \tilde{C}_\gamma  -4 N_c \text{ tr }(Q_AQ_q^2)  \tilde{C}_g  \, , \\
    \kappa =& \frac{8\pi}{\alpha_s}\frac{g_s^2 C_g}{\beta^0_\text{QCD}} \, ,\\
    \mathcal{Z} =& y_{S}  + \frac{g_s^2 C_g }{\beta^0_{\text{QCD}}} \frac{8\pi}{\alpha_s} \frac{M_q}{v} \, , \\
    \Delta^A_{ud} =& Y_P^u - Y_P^d -  16 \pi^2 \tilde{C}_g \frac{m_d-m_u}{m_d+m_u}\,,\qquad \Delta_u = 0.858(22)\,,
    \qquad\Delta_d = -0.418(22)
    \\
    \tilde{C}_{\phi\text{p}} =&  Y_P^u \Delta_u +Y_P^d \Delta_d -  16 \pi^2 \tilde{C}_g  \left[ \frac{m_u \Delta_d}{m_u + m_d} + \frac{m_d \Delta_u}{m_u + m_d} - \frac{m_\phi^2}{m_\pi^2-m_\phi^2} \frac{\Delta_u-\Delta_d}{32 \pi^2 \tilde{C}_g } \Delta^A_{ud}\right] 
    \, , \\
    \tilde{C}_{\phi\text{n}} =&  Y_P^u \Delta_d + Y_P^d \Delta_u -  16 \pi^2 \tilde{C}_g \left[ \frac{m_u \Delta_u}{m_u + m_d} + \frac{m_d \Delta_d}{m_u + m_d} + \frac{m_\phi^2}{m_\pi^2-m_\phi^2} \frac{\Delta_u-\Delta_d}{ 32 \pi^2 \tilde{C}_g } \Delta^A_{ud}\right]
    \, ,\\
    C_{\phi \text{pp}} =&  \frac{v}{m_u} y_{q,S}^u \, \sigma_u +  \frac{v}{m_d} y_{q,S}^d \, \sigma_d +  \frac{v}{m_s} y_{q,S}^s\,  \sigma_s - \frac{32\pi^2 C_g}{9} \,  (m_p - \sigma_u -\sigma_d-\sigma_s) \, ,  \\
    C_{\phi \text{nn}} =&  \frac{v}{m_d} y_{q,S}^u \, \sigma_d +  \frac{v}{m_u} y_{q,S}^d \, \sigma_u +  \frac{v}{m_s} y_{q,S}^s\,  \sigma_s - \frac{32\pi^2 C_g}{9} \,  (m_p - \sigma_u -\sigma_d-\sigma_s) \,.
\end{align}
\begin{table}[ht!]
    \centering
    \begin{tabular}{c|c|c|c|c|c}
                            & $c_\gamma$ & $y_{\ell,S}$&  $\kappa$& $\mathcal{Z}$& $C_{\phi \text{NN}}$ \\
                            \hline
      $\tilde{c}_\gamma$    & $ \tilde{c}_\gamma\,  c_\gamma $ & $ \tilde{c}_\gamma\,  y_{\ell,S}$ & $ \tilde{c}_\gamma\,  \kappa $ & $ \tilde{c}_\gamma\,  \mathcal{Z} $ & $ \tilde{c}_\gamma\, C_{\phi \text{NN}} $\\
      $y_{\ell, P}$    & $ y_{\ell, P}\,  c_\gamma $ & $ y_{\ell, P}\,  y_{\ell,S}$ & $ y_{\ell, P}\,  \kappa $ & $ y_{\ell, P}\,  \mathcal{Z} $ & $ y_{\ell, P}\,  C_{\phi \text{NN}} $\\
      $\Delta_{ud}^A$    & $ \Delta_{ud}^A\,  c_\gamma $ & $ \Delta_{ud}^A\,  y_{\ell,S}$ & $ \Delta_{ud}^A\,  \kappa $ & $ \Delta_{ud}^A\,  \mathcal{Z} $ & $ \Delta_{ud}^A\, C_{\phi \text{NN}} $\\
      $\tilde{C}_{\phi\text{N}}$    & $ \tilde{C}_{\phi\text{N}}\, c_\gamma $ & $ \tilde{C}_{\phi\text{N}}\, y_{\ell,S}$ & $ \tilde{C}_{\phi\text{N}}\, \kappa $ & $ \tilde{C}_{\phi\text{N}}\,\mathcal{Z} $ & $ \tilde{C}_{\phi\text{N}}\, C_{\phi \text{NN}} $\\
    \end{tabular}
    \caption{Jarlskog invariants emerging from the interactions in \eqref{eq:Simplest_Lag_2f_CPE} and \eqref{eq:Simplest_Lag_2f_CPO}\,.}
    \label{tab:2F_Jarlskog}
\end{table}
%

%

\section{ALP interactions for $N_f = 3$}
\label{sec:Nf3}

In this Section, we are going to generalize the previous results to the case of three flavours, $N_f = 3$.
In particular, as in the $N_f = 2$ case, we will first discuss the couplings of ALPs to mesons and then to baryons.

\subsection{ALP interactions with Mesons}

In a three-flavour setting we will allow for flavour-violating interactions in the down-sector, i.e.~the possibility of an ALP-induced mixing between the down and strange quarks occurring via the matrices
\begin{equation}
\begin{split}
 &Y_{A} = \begin{bmatrix} Y_{A}^u & 0 & 0 \\ 0 & Y_{A}^{d} & Y_{A}^{ds}\\ 0 & Y_{A}^{ds*} & Y_{A}^{s}\end{bmatrix}\qquad \text{and} \qquad  Q_A = \begin{bmatrix} q_u & 0 & 0 \\ 0 & q_{d} & q_{ds}\\ 0& q_{ds}^* & q_{s}\end{bmatrix} \, , \\
 &Y_{V} = \begin{bmatrix} 0 & 0 & 0 \\ 0 & 0 & Y_{V}^{ds}\\ 0& Y_{V}^{ds*} & 0\end{bmatrix}\quad \qquad \text{and} \qquad  Q_V = \begin{bmatrix} q_V^u & 0 & 0 \\ 0 & q_V^{d} & 0\\ 0& 0 & q_V^{s}\end{bmatrix} \, .
\end{split}
\end{equation}
 We can then expand Eqs.~\eqref{eq:ALP_Chipt_Lag_V1} and \eqref{eq:Mass_ALP_Chipt_Lag} in terms of the constituent meson fields. Up to order $\mathcal{O} (\Lambda^{-2})$ we find the following terms describing the interaction of an ALP with no more than three mesons: 
\begin{equation}
\label{eq:ALP_Chipt_Lag_3f_V1}
\begin{split}
\mathcal{L}_{\text{ALP}}^{\chi\text{pt}} &= \mathcal{L}_{\text{ALP, }\kappa}^{\chi\text{pt}}+ \mathcal{L}_{\text{ALP, }\mathcal{Z}}^{\chi\text{pt}}+ \mathcal{L}_{\text{ALP, KM}}^{\chi\text{pt}}+ \mathcal{L}_{\text{ALP, ADI}}^{\chi\text{pt}} + \mathcal{L}_{\text{ALP, VDI}}^{\chi\text{pt}} \\
&+  \mathcal{L}_{\text{ALP, MM}}^{\chi\text{pt}}+ \mathcal{L}_{\text{ALP, MI}}^{\chi\text{pt}} +   \mathcal{L}_{\text{ALP, MT}}^{\chi\text{pt}} + \mathcal{O} \left( \frac{1}{\Lambda^2}\right) \, ,
\end{split}
\end{equation}
which contains both a kinetic mixing (KM) term 
\begin{equation}
\label{eq:ALP_Chipt_Lag_3f_KM}
 \mathcal{L}_{\text{ALP, KM}}^{\chi\text{pt}} = (Y_A^u - Y_A^d)\, \frac{f_\pi}{\Lambda}\,  \partial \phi\, \partial \pi_0 + \frac{Y_A^u + Y_A^d - 2 Y_A^s}{\sqrt{3}} \, \frac{f_\pi}{\Lambda} \, \partial \eta \, \partial \phi + \sqrt{2}\, \frac{f_\pi}{\Lambda} (Y_A^{sd} \, \partial\bar{K_0} \partial \phi + Y_A^{sd*}\, \partial K_0 \partial \phi)
\end{equation}
and a mass mixing term (MM)
\begin{equation}
\label{eq:ALP_Chipt_Lag_3f_MT}
\begin{split}
\mathcal{L}_{\text{ALP, MM}}^{\chi\text{pt}} &= 2 \lambda_g m_{\pi^\pm}^2 \frac{m_u q_u - m_d q_d}{m_u + m_d}\frac{f_\pi}{\Lambda} \, \phi\, \pi_0 + \frac{2}{\sqrt{3}} \lambda_g m_{\pi^\pm}^2 \frac{m_u q_u + m_d q_d- 2 m_s q_s}{m_u + m_d}\frac{f_\pi}{\Lambda} \, \phi\, \eta_0 \\
&+ \sqrt{2} \lambda_g m_{\pi^\pm}^2 \frac{m_d+m_s}{m_u + m_d}\frac{f_\pi}{\Lambda}\, (q_{ds}\, \bar{K}_0\, \phi + q_{ds}^*\, K_0 \, \phi) \, .
\end{split}
\end{equation}
In addition to these, from the quark mass term we also extract a mass term for the ALP field (MT)
\begin{equation}
\label{eq:ALP_Chipt_Lag_3f_MM}
\mathcal{L}_{\text{ALP, MT}}^{\chi\text{pt}} = - \frac{1}{2} \frac{4 \, (\lambda_g)^2 m_{\pi^\pm}^2}{m_u + m_d} \frac{f_\pi^2}{\Lambda^2} \left[m_u q_u^2 + m_d (q_d^2 + |q_{ds}|^2) + m_s (q_s^2 + |q_{ds}|^2)\right] \phi^2
\end{equation}
and a set of ALP-meson non-derivative interactions (MI): 
\begin{equation}
\label{eq:ALP_Chipt_Lag_3f_MI}
\begin{split}
\mathcal{L}_{\text{ALP, MI}}^{\chi\text{pt}} &= - \frac{2}{3 f_\pi \Lambda} \frac{\lambda_g m_{\pi^\pm}^2}{m_u + m_d} \, \phi \, \left[\sqrt{2} \left(\bar{K}_0 K^+ \pi^- + K_0 K^- \pi^+ \right) (m_u q_u + m_d q_d + m_s q_s) - 2m_d q_d \pi_0\bar{K}_0 K_0  \right.\\
&\left. + \sqrt{3}m_sq_s \bar{K}_0 K_0\eta + m_s q_s\bar{K}_0 K_0\pi_0- \frac{4}{3\sqrt{3}} m_s q_s \eta^3 + \frac{m_d q_d}{18} \left(\sqrt{3} \eta^3 - 9 \eta^2 \pi_0 + 9 \sqrt{3} \eta \pi_0^2 - 9 \pi_0^3\right)
\right.\\
&\left.+ \frac{m_u q_u}{18} \left(\sqrt{3} \eta^3 + 9 \eta^2 \pi_0 + 9 \sqrt{3} \eta \pi_0^2 + 9 \pi_0^3\right)- K^-K^+ \left( \sqrt{3} m_s q_s \eta - (m_sq_s + 2 m_u q_u) \pi^0 \right)\right.\\
&\left.+ \pi^+ \pi^- \left( \sqrt{3} (m_u q_u + m_d q_d) \eta + (m_u q_u - m_dq_d)\pi^0 \right)\right] \\
&- \frac{\lambda_g}{ f_\pi \Lambda} \frac{m_d+m_s}{m_u + m_d}  m_{\pi^\pm}^2 \, \phi \, \left( q_{ds}\bar{K_0} + q_{ds}^*K_0\right) \left( 2 \bar{K}_0 K_0 + 2 K^+ K^- + 2 \pi^+\pi^- + \eta^2 + \pi_0^2 \right) \, .
\end{split}
\end{equation}
The other terms in the Lagrangian stem either from the term proportional to the energy-momentum tensor (subscript $\kappa$), 
\begin{equation}
\label{eq:ALP_Chipt_Lag_3f_Kappa}
\begin{split}
\mathcal{L}_{\text{ALP, }\kappa}^{\chi\text{pt}} &= \kappa \, \frac{\phi}{\Lambda} \, \left[\partial \pi_0 \partial \pi_0 + \partial \eta \partial \eta + 2 \partial \pi^+ \partial \pi^- + 2 \partial K^+ \partial K^-  + 2 \partial K_0 \partial \bar{K}_0   \right.\\
& \qquad \qquad\left.- \frac{4 m_{\pi^\pm}^2}{m_u + m_d} \left((m_d+m_s)\bar{K}_0 K_0 + (m_u+m_s)\,  K^+ K^- + (m_d+m_u) \,\pi^+ \pi^- \right.\right.\\
& \qquad \qquad \left.\left.+ (m_d+m_u+4m_s) \, \frac{\eta^2}{6}  +  (m_u-m_d) \, \frac{\eta}{\sqrt{3}} \pi_0 + (m_u + m_d) \, \frac{\pi_0^2}{2}\right)\right] \, ,
\end{split}
\end{equation}
from the scalar coupling to the matrix $\mathcal{Z}$ (subscript $\mathcal{Z}$)
\begin{equation}
\label{eq:ALP_Chipt_Lag_3f_ZMatrix}
\begin{split}
\mathcal{L}_{\text{ALP, }\mathcal{Z}}^{\chi\text{pt}} &= \frac{\phi}{\Lambda} \frac{m_{\pi^\pm}^2 \, v}{m_u + m_d} \left[ \mathcal{Z}_s \left(\bar{K}_0 K_0 + K^+ K^- + \frac{2}{3} \eta^2 \right)+ \mathcal{Z}_d \left(\bar{K}_0 K_0 + \pi^+ \pi^-+ \frac{\eta^2}{6} - \frac{\eta}{\sqrt{3}} \pi_0 + \frac{\pi_0^2}{2}\right) \right.\\ 
& \qquad \qquad \qquad \quad  \left. +\mathcal{Z}_u \left(K^+ K^- + \pi^+ \pi^-+ \frac{\eta^2}{6} + \frac{\eta}{\sqrt{3}} \pi_0 + \frac{\pi_0^2}{2}\right)\right] \, ,
\end{split}
\end{equation}
or from derivative interactions. These can be further distinguished in axial (ADI) and vectorial (VDI) derivative interactions; we find:
\begin{align}
\label{eq:ALP_Chipt_Lag_3f_ADI}
\mathcal{L}_{\text{ALP, ADI}}^{\chi\text{pt}} &= + \frac{Y_A^u- Y_A^d}{6 f_\pi \Lambda} \, \partial \phi \,  \left[ 2\sqrt{3} \partial \eta \left(\bar{K}_0 K_0 - K^+ K^-\right) - 2 \partial \pi_0 \left(\bar{K}_0 K_0 + K^+ K^- + 4 \pi^+ \pi^- \right) \right.
\nonumber\\
& \quad \left.  - \sqrt{3}\eta \left(\partial \bar{K}_0 K_0 + \bar{K}_0 \partial K_0 - D K^+ K^- - D K^- K^+\right)\right.
\nonumber\\
& \quad \left. + \pi_0 \left(\partial \bar{K}_0 K_0 + \bar{K}_0 \partial K_0 +D K^+ K^- + D K^- K^+ + 4 D \pi^+ \pi^- + 4 D\pi^- \pi^+\right) \right.
\nonumber\\
& \quad \left. + 3 \sqrt{2} \left( \partial \bar{K}_0 K^+ \pi^- +
\partial K_0 K^- \pi^+ - D K^+ \pi^-\bar{K}_0 - D K^- \pi^+ K_0\right)\right] 
\nonumber\\
& \quad + \frac{Y_A^u + Y_A^d - 2Y_A^s}{6 f_\pi \Lambda} \, \partial \phi \,  \left[- 2\sqrt{3} \partial \eta \left(\bar{K}_0 K_0 + K^+ K^-\right) + 2 \partial \pi_0 \left( \bar{K}_0 K_0 - K^+ K^-\right) \right.
\nonumber\\
& \quad \left.+ \sqrt{3}\eta \left( \partial \bar{K}_0 K_0 + \bar{K}_0 \partial K_0 + DK^+ K^- + D K^- K^+\right)
\right.\\
& \quad \left. -\pi_0 \left( \partial \bar{K}_0 K_0 + \bar{K}_0 \partial K_0 - D K^+ K^- - D K^- K^+\right)
\right.\nonumber\\
& \quad \left.+ \sqrt{2} \left( D K^+ \pi^-\bar{K}_0 +  DK^- \pi^+ K_0 +  \partial \bar{K}_0 K^+ \pi^- +  \partial K_0 K^- \pi^+ - 2 D \pi^+ K^- K_0 -2 D\pi^- K^+ \bar{K}_0 \right)\right]\nonumber\\
& \quad + \frac{1}{6 f_\pi \Lambda} \partial \phi \left[Y_A^{ds} \left(4 \sqrt{2} \partial K_0 \bar{K}_0 \bar{K}_0 + 4 \sqrt{2} D K^+ K^-\bar{K}_0 - 2DK^-(\sqrt{2}  K^+ \bar{K}_0 - 2 \sqrt{3}\pi^+ \eta) \right. \right.
\nonumber\\
& \quad \left.\left.   + 2D\pi^+ (\sqrt{3}  K^- \eta - 3 K^- \pi_0 - \sqrt{2} \pi^- \bar{K}_0) + 4 \sqrt{2} D\pi^- \pi^+ \bar{K}_0\right. \right.\nonumber\\
& \quad \left.\left.  + \partial \eta\,  (3 \sqrt{2} \bar{K}_0 \eta - \sqrt{6} \bar{K}_0 \pi_0 + 2 \sqrt{3} K^-\pi^+) + \partial \pi_0 (\bar{K}_0 \eta + \sqrt{2} \bar{K}_0 \pi_0 + 6 K^- \pi^+)\right. \right.
\nonumber\\
& \quad \left.\left.  +\sqrt{2}\,  \partial \bar{K}_0 (-4 \bar{K}_0 K_0 - 2 K^+ K^--2 \pi^+ \pi^- - 3 \eta^2 + 2 \sqrt{3} \eta \pi_0 - \pi_0^2)\right) + \text{h.c.} \right]
\nonumber
\end{align}
and 
\begin{align}
\label{eq:ALP_Chipt_Lag_3f_VDI}
\mathcal{L}_{\text{ALP, VDI}}^{\chi\text{pt}} &= + i \frac{q_V^u- q_V^d}{2 \Lambda} \, \partial \phi \,  \left[ \partial \bar{K}_0 K_0 - \partial K_0 \bar{K}_0 + DK^+ K^- - DK^- K^+ + 2 D\pi^+ \pi^- - 2D\pi^- \pi^+\right]
\nonumber\\
& \quad  - i \frac{q_V^u + q_V^d - 2 q_V^s}{2 \Lambda} \, \partial \phi \, \left[\partial \bar{K}_0 K_0 - \partial K_0 \bar{K}_0 + DK^- K^+ - DK^+ K^-\right]\\
& \quad +i \left[\frac{Y_V^{ds}}{2\Lambda}\,  \partial \phi \, \left( - \sqrt{6} \partial \eta \bar{K}_0 + \sqrt{6} \partial \bar{K}_0 \eta + \sqrt{2} \partial \pi_0 \bar{K}_0 - \sqrt{2} \partial \bar{K}_0 \pi_0 + 2 DK^- \pi^+ - 2 D\pi^+ K^- \right)  -\text{h.c.}\right] \, .
\nonumber
\end{align}
The two terms in Eqs.~\eqref{eq:ALP_Chipt_Lag_3f_KM} and \eqref{eq:ALP_Chipt_Lag_3f_MM} are responsible for both kinetic and mass mixing between the ALP field and the neutral mesons $\pi_0, \eta, \bar{K}_0, K_0$. These fields can be put in a canonical form by repeating the procedure outlined in Section \ref{sec:ALPINT2F} for the full $5 \times 5$ mixing matrices. 

However, this does not allow us to obtain analytical results for the mixing, and a better strategy is to block-diagonalize the various mixing terms: in this case, as we will see, the results are analytical and start differing from the exact ones at order $\mathcal{O} (\Lambda^{-2})$ in the mixing terms and at order $\mathcal{O}(\Lambda^{-4})$ in the mass terms. 

First of all we define the small parameter $\xi = f_\pi /\Lambda $ and
\begin{equation}
A = Y^u_A - Y^d_A \, ,  \qquad B = \frac{Y_A^u + Y_A^d - 2 Y_A^s}{\sqrt{3}} \, , \qquad C = 2 \, \text{Re} Y_A^{ds} \, , \qquad G = -2 \, \text{Im} Y_A^{ds} \, , 
\end{equation}
as well as 
\begin{equation}
\begin{split}
 &\alpha' = 2\, (m_u q_u - m_d q_d) \lambda_g \frac{m_{\pi^{\pm}}^2}{m_u + m_d} \, , \qquad \beta' = \frac{2}{\sqrt{3}} \, (m_u q_u + m_d q_d - 2m_s q_s) \lambda_g \frac{m_{\pi^{\pm}}^2}{m_u + m_d} \, , \qquad\\
 & \gamma' = 2 \, \text{Re} (q_{ds}) (m_d + m_s) \lambda_g \frac{m_{\pi^{\pm}}^2}{m_u + m_d} \, , \qquad \delta' = -2\, \text{Im} (q_{ds}) (m_d + m_s) \lambda_g \frac{m_{\pi^{\pm}}^2}{m_u + m_d} \, .
\end{split}
\end{equation}
Then we can start diagonalizing the sub-sector responsible for the mixing between the ALP and the neutral pion. According to our results in Eq.~\eqref{eq:Physical_Fields_2f} we obtain
\begin{equation}
\label{eq:Physical_Fields_3f_ALP_Pi0}
\begin{split}
    &\phi' = \phi + \xi A \frac{m_{\pi_0}^2 + (\alpha'/A)}{m_{\pi_0}^2-m_\phi^2} \pi_0 \, , \\
    &\pi_{0,\text{ph}} = \pi_0 - \xi A \frac{m_\phi^2 + (\alpha'/A)}{m_{\pi_0}^2-m_\phi^2} \phi \, , 
\end{split}
\end{equation}
which can be inverted, yielding
\begin{equation}
\label{eq:Physical_Fields_3f_ALP_Pi0_Inverse}
\begin{split}
    &\phi = \phi' - \xi A \frac{m_{\pi_0}^2 + (\alpha'/A)}{m_{\pi_0}^2-m_\phi^2} \pi_0 \, , \\
    &\pi_0 = \pi_{0,\text{ph}} + \xi A \frac{m_\phi^2 + (\alpha'/A)}{m_{\pi_0}^2-m_\phi^2} \phi' \, .
\end{split}
\end{equation}
Now it is apparent what we anticipated previously: the only modifications to the mixing of ALPs with the other mesons start appearing at order $\mathcal{O}(\xi^2) = \mathcal{O}(\Lambda^{-2})$ and can be safely neglected. For instance, the mass mixing term between the ALP and the $\eta$ meson is shifted according to 
\begin{equation}
    \beta' \xi \,  \phi\,  \eta = \beta' \xi \, \left( \phi' - \xi A \frac{m_{\pi_0}^2 + (\alpha'/A)}{m_{\pi_0}^2-m_\phi^2} \pi_0 \right) \, \eta = \beta' \xi \, \phi' \, \eta + \mathcal{O}(\xi^2) \, .
\end{equation}
The ALP-induced mixing in the $\pi_0-\eta$ sector thus emerges only at order $\xi^2$, yielding mass corrections of order $\xi^4$ and higher; therefore, we can safely neglect this kind of second-order effects in our analysis. The reasoning we have followed here is completely general and applies as well as to the kinetic mixing term, besides being valid of course also for the other possible mixing effects with the remaining neutral mesons. 

We repeat this procedure also for the remaining fields, finding in the end that, up to $\mathcal{O}(\Lambda^{-2})$
\begin{equation}
\begin{split}
    &\phi_{\text{ph}} = \phi +  \xi A \frac{m_{\pi_0}^2 + (\alpha'/A)}{m_{\pi_0}^2-m_\phi^2} \pi_0  + \xi B \frac{m_\eta^2 + (\beta'/B)}{m_\eta^2-m_\phi^2} \eta + \xi C \frac{m_{K_0}^2 + (\gamma'/C)}{m_{K_0}^2-m_\phi^2} \pi_6 + \xi G \frac{m_{K_0}^2 + (\delta'/G)}{m_{K_0}^2-m_\phi^2} \pi_7 \, , \\
    &\pi_0 = \pi_{0,\text{ph}} + \xi A \frac{m_{\phi}^2 + (\alpha'/A)}{m_{\pi_0}^2-m_\phi^2} \phi_{\text{ph}} \, , \\
    &\eta = \eta_{\text{ph}} + \xi B \frac{m_\phi^2 + (\beta'/B)}{m_\eta^2-m_\phi^2} \phi_{\text{ph}} \, , \\
    &\pi_6 = \pi_{6,\text{ph}} + \xi C \frac{m_{\phi}^2 + (\gamma'/C)}{m_{K_0}^2-m_\phi^2} \phi_{\text{ph}} \, , \\
    &\pi_7 = \pi_{7,\text{ph}} + \xi G \frac{m_{\phi}^2 + (\delta'/G)}{m_{K_0}^2-m_\phi^2} \phi_{\text{ph}} \, .
\end{split}
\end{equation}
These mixing effects are then to be taken into account into the meson-meson interaction terms stemming from both the kinetic and the mass one, as well as into the anomalous coupling of the $\pi_0$ and the $\eta$ mesons with two photons. 
Recalling that these interactions read
\begin{equation}
\begin{split}
\mathcal{L}_{\pi \text{KT}} &\supset \frac{1}{12f_\pi^2} \left[ 2\partial \bar{K}_0 \partial \bar{K}_0 K_0 K_0 + 2 \partial K_0 \partial K_0 \bar{K}_0 \bar{K}_0 - \partial \pi_0 \partial \pi_0 \left(\bar{K}_0 K_0 + K^+ K^- + 4 \pi^+ \pi^-\right)\right. \\
& \quad \left.- \partial \bar{K}_0 \partial K_0 \left( 4 \bar{K}_0 K_0 + 2 K^+ K^- + 3 \eta^2 - 2 \sqrt{3} \eta \pi_0 + \pi_0^2 + 2 \pi^+ \pi^-\right)- 3\partial \eta \partial \eta \left( \bar{K}_0 K_0 + K^+ K^-\right) \right.\\
& \quad \left. -\partial \bar{K}_0\left(-3 \partial \eta \eta K_0+ \sqrt{3} \partial \pi_0 \eta K_0 + \sqrt{3} \partial \eta \pi_0 K_0 - \partial \pi_0 \pi_0 K_0 + \sqrt{6} \partial \eta K^+ K^- + 3\partial \pi_0 K^+ K^-\right) \right.\\
& \quad \left.- \partial K_0\left(-3 \partial \eta \eta \bar{K}_0+ \sqrt{3} \partial \pi_0 \eta \bar{K}_0 + \sqrt{3} \partial \eta \pi_0 \bar{K}_0 - \partial \pi_0 \pi_0 \bar{K}_0 + \sqrt{6} \partial \eta K^+ K^-  + 3\partial \pi_0 K^+ K^-\right)   \right]   
\end{split}
\end{equation}
 and 
 \begin{equation}
 \begin{split}
\mathcal{L}_{\pi \text{MT}} &\supset \frac{m_{\pi^\pm}^2}{m_u+m_d}\frac{1}{f_\pi^2} \left[\frac{m_u + m_d + 16 m_s}{216} \eta^4 + \frac{(m_u-m_d)}{18\sqrt{3}} \eta^3 \pi_0  + \frac{\eta^2}{12} \left(\bar{K}_0 K_0 (m_d + 3 m_s) + K^+ K^- (m_u + 3 m_s) \right. \right.\\
& \quad \left. \left.+ (\pi_0^2 + 2 \pi^+\pi^-) (m_u-m_s)\right) + \frac{\eta}{12 \sqrt{3}} \left( 2(m_u-m_d) \pi_0^3 + \sqrt{2}(m_u + m_d -2 m_s) (\bar{K}_0 K^+ \pi^- + K_0 K^- \pi^+) \right. \right.\\
& \quad \left. \left. + 2 \pi_0 \left((m_s-m_d)\bar{K}_0 K_0 + (m_u-m_s)K^+ K^- + 2 (m_u-m_d)\pi^+ \pi^-\right)\right) \right.\\
& \left. \quad + \frac{m_u + m_d}{24} \pi_0^4 +\frac{\pi_0^2}{12} \left( \bar{K}_0 K_0 (3m_d + m_s) + K^+ K^- (3 m_u +m_s) + 2 \pi^+ \pi^-(m_u + m_d)\right) \right.\\
& \quad \left.+ \frac{m_u-m_d}{6 \sqrt{2}}\, \pi_0 \,\left(\bar{K}_0 K^+ \pi^- + K_0 K^- \pi^+\right) + \frac{\bar{K}_0 K_0}{6} \left(K^+ K^- (m_u + m_d + 2m_s) + \pi^+ \pi^- (m_u + 2m_d + m_s)\right) \right] \, ,
 \end{split}
 \end{equation}
 one can perform the needed shifts in the neutral meson fields. 
For instance, one finds 
\begin{equation}
\begin{split}
\mathcal{L}_{\pi \text{KT+MT}}^A &= \xi A \frac{m_\phi^2 + \alpha'/A}{m_{\pi_0}^2-m_\phi^2} \frac{1}{12 f_\pi^2} 
\left[ -2 \partial \phi \partial \pi_0 \left(\bar{K}_0 K_0 + K^+ K^- + 4 \pi^+ \pi^-\right) - \partial \bar{K}_0 \partial K_0 \left(-2\sqrt{3} \eta \phi + 2 \pi_0 \phi\right)\right.\\
& \quad \left. - \partial \bar{K}_0 \left(\sqrt{3}\partial \phi \eta K_0 + \sqrt{3} \partial \eta \phi K_0 -  \partial \phi \pi_0 K_0 - \partial \pi_0 \phi K_0 + 3 \partial \phi K^+ K^-\right)\right.\\
& \quad \left.
- \partial K_0 \left(\sqrt{3}\partial \phi \eta \bar{K}_0 + \sqrt{3} \partial \eta \phi \bar{K}_0 -  \partial \phi \pi_0 \bar{K}_0 - \partial \pi_0 \phi \bar{K}_0 + 3 \partial \phi K^+ K^-\right) \right]+\\
& \quad\xi A \frac{m_\phi^2 + \alpha'/A}{m_{\pi_0}^2-m_\phi^2} \frac{m_{\pi^\pm}^2}{m_u + m_d}\frac{1}{f_\pi^2} \left[\frac{m_u-m_d}{18 \sqrt{3}} \eta^3 \phi  +  \frac{\eta^2}{6}\phi \pi_0(m_u-m_s) + \frac{\eta}{2 \sqrt{3}} \pi_0^2 \phi (m_u-m_d)\right.\\
& \quad\left. + \frac{\eta}{6\sqrt{3}} \phi \left((m_s-m_d) \bar{K}_0 K_0 + (m_u-m_s)K^+K^- + 2 (m_u-m_d)\pi^+\pi^-\right) \right.\\
& \left.\quad  + \frac{m_u + m_d}{6} \phi \pi_0^3 + \frac{\pi_0}{6}\phi \left( \bar{K}_0 K_0 (3m_d + m_s) + K^+ K^- (3 m_u +m_s) + 2 \pi^+ \pi^-(m_u + m_d)\right)\right]\\
& \quad +  \xi A \frac{m_\phi^2 + \alpha'/A}{m_{\pi_0}^2-m_\phi^2} \frac{e^2 }{16\pi^2f_\pi} \phi F^{\mu\nu} \tilde{F}_{\mu\nu} \, , 
\end{split}
\end{equation}
\begin{equation}
\label{eq:Eta_ALP_Mixing_Interactions}
\begin{split}
\mathcal{L}_{\eta \text{KT+MT}}^B &= \xi B \frac{m_\phi^2 + \beta'/B}{m_{\eta}^2-m_\phi^2} \frac{1}{12 f_\pi^2} \left[  - 6 \partial \phi \partial\eta (\bar{K}_0 K_0 + K^+ K^-)\right.\\
& \quad \left.-\partial \bar{K}_0 \left(- 3 \partial \phi \eta K_0 - 3 \partial \eta \phi K_0 +  \sqrt{3} \partial \pi_0 \phi K_0 + \sqrt{3} \partial \phi \pi_0 K_0 + \sqrt{6} \partial \phi K^+ K^-\right)- \partial \bar{K}_0 \partial K_0 \left(6 \eta \phi - 2 \sqrt{3} \phi \pi_0 \right)\right.\\
& \quad \left.-\partial K_0 \left(- 3 \partial \phi \eta \bar{K}_0 - 3 \partial \eta \phi \bar{K}_0 +  \sqrt{3} \partial \pi_0 \phi \bar{K}_0 + \sqrt{3} \partial \phi \pi_0 \bar{K}_0 + \sqrt{6} \partial \phi K^+ K^-\right)\right] + \\
& \quad\xi B \frac{m_\phi^2 + \beta'/B}{m_{\eta}^2-m_\phi^2} \frac{m_{\pi^\pm}^2}{m_u + m_d}\frac{1}{f_\pi^2}  \left[\frac{m_u + m_d + 16 m_s}{54} \eta^3 \phi + \frac{(m_u-m_d)}{6\sqrt{3}} \eta^2 \phi \pi_0  + \frac{\eta \phi }{6} \left(\bar{K}_0 K_0 (m_d + 3 m_s)\right. \right.\\
& \quad \left. \left. + K^+ K^- (m_u + 3 m_s) + (\pi_0^2 + 2 \pi^+\pi^-) (m_u-m_s)\right) + \frac{\phi}{12 \sqrt{3}} \left( 2(m_u-m_d) \pi_0^3\right. \right.\\
& \quad \left. \left. + \sqrt{2}(m_u + m_d -2 m_s) (\bar{K}_0 K^+ \pi^- + K_0 K^- \pi^+)  + 2 \pi_0 \left((m_s-m_d)\bar{K}_0 K_0 + (m_u-m_s)K^+ K^- \right. \right.\right.\\
& \quad \left.\left. \left.+ 2 (m_u-m_d)\pi^+ \pi^-\right)\right)  \right]+  \xi B \frac{m_\phi^2 + \beta'/B}{m_{\eta}^2-m_\phi^2} \frac{e^2 }{16\pi^2f_\pi} \phi F^{\mu\nu} \tilde{F}_{\mu\nu} \, ,
\end{split}
\end{equation}
and similarly for the mixing between the ALP field and the neutral kaons. We do not report here the explicit expressions for such a mixing effect because, as it will be clear soon, the charges $q_{ds}$ can be always chosen in such a way to eliminate it completely.
Indeed, a convenient choice consists in minimizing the possible mixing effects in the mesonic sector. This amounts, for instance, to requiring that
\begin{equation}
\label{eq:Minimal_Mixing_Choice}
\frac{\alpha'}{A} = -m_\phi^2 \, , \qquad \frac{\gamma'}{C} = -m_\phi^2 \, , \qquad \frac{\delta'}{G} = -m_\phi^2 \, , 
\end{equation}
which corresponds to the choice 
\begin{equation}
\label{eq:Explicit_Charge_Assignment}
\begin{split}
q_u &= \frac{1}{2} \frac{m_d m_s}{m_u m_d + m_u m_s + m_d m_s} - \frac{m_\phi^2}{m_{\pi_\pm}^2-m_\phi^2} \frac{\Delta_A}{2 \lambda_g} \, , \\
q_d &= \frac{1}{2} \frac{m_u m_s}{m_u m_d + m_u m_s + m_d m_s} + \frac{m_\phi^2}{m_{\pi^\pm}^2-m_\phi^2} \frac{\Delta_A}{2 \lambda_g} \, , \\
q_s &= \frac{1}{2} \frac{m_u m_d}{m_u m_d + m_u m_s + m_d m_s} \, ,  \\
q_{ds} &= -\frac{Y_{P}^{ds}}{\lambda_g}\frac{m_\phi^2 (m_u +m_d)}{(m_d+m_s) m_{\pi^{\pm}}^2 - m_\phi^2(m_u+m_d)} \, , 
\end{split}
\end{equation}
where, also for future reference, we define
\begin{equation}
    \begin{split}
    \Delta_A &= Y_P^u  -Y_P^d - \frac{\lambda_g}{2} \frac{m_s(m_d-m_u)}{m_u m_d + m_u m_s + m_d m_s} = \frac{m_{\pi_\pm}^2 - m_\phi^2}{m_{\pi_\pm}^2} (Y_A^u-Y_A^d) \, , \\
    \Omega_A &= Y_P^u  +Y_P^d -2Y_P^s - \frac{\lambda_g}{2}\frac{m_s(m_u+m_d)-2 m_u m_d}{m_u m_d + m_u m_s + m_d m_s}=  (Y_A^u+Y_A^d-2 Y_A^s) \, , \\
    \Xi_A &=  Y_{P}^{ds}\frac{m_{\pi^\pm}^2 (m_d+m_s)}{(m_d+m_s) m_{\pi^{\pm}}^2 - m_\phi^2(m_u+m_d)} \, , \\
    \Gamma_A &= Y_P^u + Y_P^d + Y_P^s - \frac{\lambda_g}{2} = (Y_A^u+Y_A^d+ Y_A^s) \, .
    \end{split} 
\end{equation}

It should be noticed, though, that this choice does not remove the mixing between the ALP and both the $\pi_0$ and $\eta$ fields. The removal of mixing between the ALP and both $\pi_0$ and $\eta$ can be achieved via the choice: 
\begin{align}
q_u &= \frac{1}{2D_M} \left[m_d m_s  - Y_A^u (m_d + m_s)(m_d + m_u)\frac{m_\phi^2}{m_{\pi_0}^2}+ Y_A^d m_s(m_d + m_u)\frac{m_\phi^2}{m_{\pi_0}^2} + Y_A^s m_d(m_d + m_u)\frac{m_\phi^2}{m_{\pi_0}^2} \right]\,, \\
q_d &= \frac{1}{2D_M} \left[m_u m_s  + Y_A^u m_s(m_d + m_u)\frac{m_\phi^2}{m_{\pi_0}^2}- Y_A^d (m_u + m_s)(m_d + m_u)\frac{m_\phi^2}{m_{\pi_0}^2} + Y_A^s m_u (m_d + m_u)\frac{m_\phi^2}{m_{\pi_0}^2} \right]\,, \\
q_s &= \frac{1}{2D_M} \left[m_u m_d  + Y_A^u m_d(m_d + m_u)\frac{m_\phi^2}{m_{\pi_0}^2}+ Y_A^d m_u(m_d + m_u)\frac{m_\phi^2}{m_{\pi_0}^2} - Y_A^s (m_d + m_u)^2\frac{m_\phi^2}{m_{\pi_0}^2} \right]\,, 
\end{align}
where $D_M = m_u m_s + m_d m_s + m_u m_d$. The price to be paid for this choice amounts to more complicated expressions in intermediate steps. Since most of the applications of our Lagrangian do not involve directly the $\eta$ meson, we find it better to employ the charge assignment in Eq.~\eqref{eq:Explicit_Charge_Assignment}, which results in more compact expressions for the ALP couplings to pions. \\

In any case, with our choice \eqref{eq:Explicit_Charge_Assignment}, the ALP acquires a mass term given by 
\begin{equation}
\label{eq:ALP_Mass_3f}
    m_{\text{ph}}^2 = m_\phi^2 - \xi^2 B^2 \frac{(m_\phi^2+ \beta'/B)^2}{m_{\eta}^2 - m_\phi^2} = m_\phi^2 - \frac{\Omega_A^2}{3}\frac{f_\pi^2}{\Lambda^2}\frac{M_\phi^4}{m_\eta^2- M_\phi^2} \left[1 + \frac{\Delta_A}{\Omega_A} \frac{m_d-m_u}{m_u+m_d} \frac{m_{\pi^\pm}^2}{m_{\pi^\pm}^2- M_\phi^2}\right]^2 + \mathcal{O} \left(\frac{1}{\Lambda^4}\right) \, , 
\end{equation}
where
\begin{equation}
\label{eq:ALP_Mass_3f_Part2}
\begin{split}
    m_\phi^2 = M_\phi^2 + m_{\pi^\pm}^2\frac{f_\pi^2}{\Lambda^2} &\left[(\lambda_g)^2 \, \frac{m_u m_s m_d}{m_u+m_d}\frac{1}{m_u m_d + m_u m_s + m_d m_s} + 4 \,  |\Xi_A|^2\frac{M_\phi^4}{m_{\pi^\pm}^4} \frac{m_d+m_u}{m_s+m_d} \right.\\
    & \left. \quad + \Delta_A^2\frac{M_\phi^4}{(m_{\pi^{\pm}}^2-M_\phi^2)^2}\right] +  \mathcal{O} \left(\frac{1}{\Lambda^4}\right) \, , 
\end{split}
\end{equation}
while the interaction terms (MI and ADI) now read
\begin{align}
\label{eq:ALP_Chipt_Lag_3f_MI_v2}
\mathcal{L}_{\text{ALP, MI}}^{\chi\text{pt}} &= - \frac{1}{3 f_\pi \Lambda} \frac{\lambda_g m_{\pi^\pm}^2}{m_u + m_d} \, \phi \, \left[\sqrt{2} \left(\bar{K}_0 K^+ \pi^- + K_0 K^- \pi^+ \right) \left(3 \tilde{m} + \tilde{\Delta}_A(m_d-m_u)\right) - 2 \left(\tilde{m} + m_d \tilde{\Delta}_A\right) \pi_0\bar{K}_0 K_0  \right.
\nonumber\\
& \quad \left. + \sqrt{3} \tilde{m} \bar{K}_0 K_0\eta + \tilde{m}\bar{K}_0 K_0\pi_0- \frac{4}{3\sqrt{3}} \tilde{m} \eta^3 + \frac{1}{18}\left(\tilde{m} + m_d \tilde{\Delta}_A\right) \left(\sqrt{3} \eta^3 - 9 \eta^2 \pi_0 + 9 \sqrt{3} \eta \pi_0^2 - 9 \pi_0^3\right)
\right.\nonumber\\
& \quad \left.+ \frac{1}{18} \left(\tilde{m} - m_u \tilde{\Delta}_A\right)\left(\sqrt{3} \eta^3 + 9 \eta^2 \pi_0 + 9 \sqrt{3} \eta \pi_0^2 + 9 \pi_0^3\right)- K^-K^+ \left( \sqrt{3} \tilde{m} \eta - (3\tilde{m} -2 \tilde{\Delta}_A m_u) \pi^0 \right)\right.\nonumber\\
& \quad \left.+ \pi^+ \pi^- \left( \sqrt{3} (2 \tilde{m} + (m_d-m_u)\tilde{\Delta}_A) \eta + (m_u + m_d)\tilde{\Delta}_A\pi^0 \right)\right] \nonumber\\
&  \quad + \frac{M_\phi^2}{ f_\pi \Lambda} \, \phi \, \left( \Xi_A\bar{K_0} + \Xi_A^*K_0\right) \left( 2 \bar{K}_0 K_0 + 2 K^+ K^- + 2 \pi^+\pi^- + \eta^2 + \pi_0^2 \right) 
\end{align}
and
\begin{align}
\label{eq:ALP_Chipt_Lag_3f_ADI_v2}
\mathcal{L}_{\text{ALP, ADI}}^{\chi\text{pt}} &= + \frac{1}{6 f_\pi \Lambda} \frac{m_{\pi^\pm}^2}{m_{\pi^\pm}^2-m_\phi^2} \Delta_A \, \partial \phi \,  \left[ 2\sqrt{3} \partial \eta \left(\bar{K}_0 K_0 - K^+ K^-\right) - 2 \partial \pi_0 \left(\bar{K}_0 K_0 + K^+ K^- + 4 \pi^+ \pi^- \right) \right.
\nonumber\\
& \quad \left.  - \sqrt{3}\eta \left(\partial \bar{K}_0 K_0 + \bar{K}_0 \partial K_0 - D K^+ K^- - D K^- K^+\right)\right.
\nonumber\\
& \quad \left. + \pi_0 \left(\partial \bar{K}_0 K_0 + \bar{K}_0 \partial K_0 +D K^+ K^- + D K^- K^+ + 4 D \pi^+ \pi^- + 4 D\pi^- \pi^+\right) \right.
\nonumber\\
& \quad \left. + 3 \sqrt{2} 
\left( \partial \bar{K}_0 K^+ \pi^- + \partial K_0 K^- \pi^+ - D K^+ \pi^-\bar{K}_0 - D K^- \pi^+ K_0\right)\right] 
\nonumber\\
& \quad + \frac{\Omega_A}{6 f_\pi \Lambda} \, \partial \phi \left[- 2\sqrt{3} \partial \eta \left(\bar{K}_0 K_0 + K^+ K^-\right) + 2 \partial \pi_0 \left( \bar{K}_0 K_0 - K^+ K^-\right) \right.
\nonumber\\
& \quad \left.+ \sqrt{3}\eta \left( \partial \bar{K}_0 K_0 + \bar{K}_0 \partial K_0 + DK^+ K^- + D K^- K^+\right)
\right.\\
& \quad \left. -\pi_0 \left( \partial \bar{K}_0 K_0 + \bar{K}_0 \partial K_0 - D K^+ K^- + D K^- K^+\right)
\right.\nonumber\\
& \quad \left.+ \sqrt{2} \left( D K^+ \pi^-\bar{K}_0 +  DK^- \pi^+ K_0 +  \partial \bar{K}_0 K^+ \pi^- +  \partial K_0 K^- \pi^+ - 2 D \pi^+ K^- K_0 -2 D\pi^- K^+ \bar{K}_0 \right)\right]
\nonumber\\
& \quad + \frac{1}{6 f_\pi \Lambda} \partial \phi \left[\Xi_A \left(4 \sqrt{2} \partial K_0 \bar{K}_0 \bar{K}_0 + 4 \sqrt{2} D K^+ K^-\bar{K}_0 - 2DK^-(\sqrt{2}  K^+ \bar{K}_0 - 2 \sqrt{3}\pi^+ \eta) \right. \right.
\nonumber\\
& \quad \left.\left.   + 2D\pi^+ (\sqrt{3}  K^- \eta - 3 K^- \pi_0 - \sqrt{2} \pi^- \bar{K}_0) + 4 \sqrt{2} D\pi^- \pi^+ \bar{K}_0\right. \right.
\nonumber\\
& \quad \left.\left.  + \partial \eta\,  (3 \sqrt{2} \bar{K}_0 \eta - \sqrt{6} \bar{K}_0 \pi_0 + 2 \sqrt{3} K^-\pi^+) + \partial \pi_0 (\bar{K}_0 \eta + \sqrt{2} \bar{K}_0 \pi_0 + 6 K^- \pi^+)\right. \right.
\nonumber\\
& \quad \left.\left.  +\sqrt{2}\,  \partial \bar{K}_0 (-4 \bar{K}_0 K_0 - 2 K^+ K^--2 \pi^+ \pi^- - 3 \eta^2 + 2 \sqrt{3} \eta \pi_0 - \pi_0^2)\right) + \text{h.c.} \right] 
\nonumber\, , 
\end{align}
where we have conveniently defined 
\begin{equation}
\tilde{m} \equiv \frac{m_u m_d m_s}{m_u m_d + m_u m_s + m_d m_s} \qquad \text{and} \qquad \tilde{\Delta}_A = \frac{m_\phi^2}{m_{\pi^\pm}^2- m_\phi^2} \frac{\Delta_A}{\lambda_g} \, .
\end{equation}
With the charge assignment in Eq.~\eqref{eq:Explicit_Charge_Assignment} then one finds that all of the interactions between one ALP and at most three pseudoscalar mesons at order $\mathcal{O}(\Lambda^{-2})$ are given by the sum of Eqs.~\eqref{eq:ALP_Chipt_Lag_3f_Kappa}, \eqref{eq:ALP_Chipt_Lag_3f_ZMatrix}, \eqref{eq:ALP_Chipt_Lag_3f_VDI}, \eqref{eq:ALP_Chipt_Lag_3f_MI_v2} and \eqref{eq:ALP_Chipt_Lag_3f_ADI_v2}, with an ALP mass given by Eqs.~\eqref{eq:ALP_Mass_3f} and \eqref{eq:ALP_Mass_3f_Part2}. 

These have then to be supplemented with the explicit mixing term \eqref{eq:Eta_ALP_Mixing_Interactions}, where $B$ and $\beta'$ are specialized to our explicit choice of charges in 
Eq.~\eqref{eq:Explicit_Charge_Assignment}: 
\begin{align}
\label{eq:Eta_ALP_Mixing_Interactions_Our_Charges}
\mathcal{L}_{\eta \text{KT+MT}}^{B} &= \frac{ \Omega_A}{\sqrt{3}} \frac{f_\pi}{\Lambda} \frac{m_\phi^2 }{m_{\eta}^2-m_\phi^2} \left(1 + \frac{\Delta_A}{\Omega_A} \frac{m_d-m_u}{m_d+m_u} \frac{m_{\pi^\pm}^2}{m_{\pi^\pm}^2- m_\phi^2}\right)  \left\{\frac{1}{12 f_\pi^2} \left[ - \partial \bar{K}_0 \partial K_0 \left(6 \eta \phi - 2 \sqrt{3} \phi \pi_0 \right)\right. \right.
\nonumber\\
& \quad \left.\left.-\partial \bar{K}_0 \left(- 3 \partial \phi \eta K_0 - 3 \partial \eta \phi K_0 +  \sqrt{3} \partial \pi_0 \phi K_0 + \sqrt{3} \partial \phi \pi_0 K_0 + \sqrt{6} \partial \phi K^+ K^-\right)\right.\right.
\nonumber\\
& \quad \left.\left.-\partial K_0 \left(- 3 \partial \phi \eta \bar{K}_0- 3 \partial \eta \phi \bar{K}_0 +  \sqrt{3} \partial \pi_0 \phi \bar{K}_0 + \sqrt{3} \partial \phi \pi_0 \bar{K}_0 + \sqrt{6} \partial \phi K^+ K^-\right)\right.\right.
\nonumber\\
&\left.\left. \quad- 6 \partial \phi \partial \eta (\bar{K}_0 K_0 + K^+ K^-)\right] \right.\\
& \quad \left.+ \frac{m_{\pi^\pm}^2}{m_u + m_d}\frac{1}{f_\pi^2}  \left[\frac{m_u + m_d + 16 m_s}{54} \eta^3 \phi + \frac{(m_u-m_d)}{6\sqrt{3}} \eta^2 \phi \pi_0  + \frac{\eta \phi }{6} \left(\bar{K}_0 K_0 (m_d + 3 m_s)\right. \right.\right.
\nonumber\\
& \quad \left.\left. \left. + K^+ K^- (m_u + 3 m_s) + (\pi_0^2 + 2 \pi^+\pi^-) (m_u-m_s)\right) + \frac{\phi}{12 \sqrt{3}} \left( 2(m_u-m_d) \pi_0^3\right. \right.\right.
\nonumber\\
& \quad \left. \left. \left. + \sqrt{2}(m_u + m_d -2 m_s) (\bar{K}_0 K^+ \pi^- + K_0 K^- \pi^+)  + 2 \pi_0 \left((m_s-m_d)\bar{K}_0 K_0 + (m_u-m_s)K^+ K^- \right. \right.\right.\right.
\nonumber\\
& \quad \left. \left.\left. \left.+ 2 (m_u-m_d)\pi^+ \pi^-\right)\right)  \right]+ \frac{e^2 }{16\pi^2f_\pi} \phi F^{\mu\nu} \tilde{F}_{\mu\nu} \right\} \, .
\nonumber
\end{align}
An interesting feature of the final Lagrangian is that, as it was the case for the results in a two-flavour setting, one is always free to choose the diagonal vectorial charges in such a way to get rid of the corresponding term from the Lagrangian. However, in a three-flavour setting the flavour-changing vectorial interactions are physical and the corresponding effects are to be considered in the last line of Eq.~\eqref{eq:ALP_Chipt_Lag_3f_VDI}. 

Of course, one could implement a rotation of quark fields via a non-diagonal vectorial charge matrix $Q_V$ in Eq.~\eqref{eq:Quark_Field_Redefinition}. This could be used in order to rotate away indeed the last line of \eqref{eq:ALP_Chipt_Lag_3f_VDI} at the price of introducing an additional source of mixing via a mass term of the kind 
\begin{equation}
\begin{split}
\begin{split}
i \frac{\phi}{\Lambda} \frac{f_\pi^2}{2} B_0 \text{Tr} \left[\left[M_q, Q_V\right] (\Sigma^\dagger - \Sigma)\right] - \frac{1}{2} \frac{\phi^2}{\Lambda^2}\frac{f_\pi^2}{2} B_0 \text{Tr} \left[\left[Q_V,\left[ Q_V, M_q\right]\right] (\Sigma^\dagger + \Sigma)\right] + \mathcal{O} \left( \frac{1}{\Lambda^2}\right) \, , 
\end{split}
\end{split}
\end{equation}
which would no longer vanish, since a non-diagonal $Q_V$ does not commute with $M_q$. Note, also, that introducing a non-diagonal $Q_V$ matrix would  modify the value of the ALP mass with an extra contribution, in a way that is completely analogous to the $Q_A$ case (but with a different structure of commutators and anticommutators). 
Since we do not see any reason to prefer this second approach over the one we have so far followed, we stick to our choice and choose $Q_V$ to be diagonal.

\subsection{ALP interactions with Baryons}

We have already discussed the three possible terms from which an interaction between ALPs and nucleons arise. These very terms are of course responsible for the couplings of an ALP to baryons in a generalized three-flavour setting.
Exploring such kinds of interactions will be the subject of this Section.
First of all, we stress once again the fact that the axial and vectorial currents in \eqref{eq:QCD_Scale_ALP_Lag_V2} will in general cointain also a baryonic component of the form discussed in \eqref{eq:Baryonic_Currents}. 
These ought to be supplemented with the $U(1)_A$ axial current 
in Eq.~(\ref{eq:jABp}), resulting in the following Lagrangian describing the interactions of an ALP and at most two baryons at a time: 
\begin{equation}
\label{eq:BarDerALP_1}
\begin{split}
\mathcal{L}_{\text{ALP}}^{\partial B} &= \frac{\partial_\mu \phi}{\Lambda} \left[\text{Tr} (Y_A \lambda_a) (j_{A, a}^\mu)_{\text{B}} + \frac{1}{3} \text{Tr}(Y_A) (j_{A}^\mu)'_{\text{B}} + \text{Tr} (Y_V \lambda_a) (j_{V, a}^\mu)_{\text{B}} \right]\\
&= \frac{\partial_\mu \phi}{\Lambda} \left[\text{Tr} (Y_A \lambda_a) \left[ D \, \text{Tr}(\bar{B}_v \gamma^\mu \gamma_5 \{T_a, B_v\})+ F\, \text{Tr}(\bar{B}_v \gamma^\mu \gamma_5 \left[T_a, B_v\right])\right] + \frac{1}{3} \text{Tr}(Y_A)\, S \, \text{Tr}(\bar{B}_v \gamma^\mu \gamma_5  B_v) \right.\\
& \qquad \qquad \left.- \text{Tr} (Y_V \lambda_a) \text{Tr}(\bar{B}_v \gamma^\mu \left[T_a, B_v\right])\right] \, , 
\end{split}
\end{equation}
where $a = {3, 8, 6, 7}$. Explicitly, with the choice in Eq.~\eqref{eq:Minimal_Mixing_Choice} we obtain 
\begin{align}
\label{eq:BarDerALP_v1}
\mathcal{L}_{\text{ALP}}^{\partial B} &= \frac{\partial_\mu \phi}{3\Lambda} \Gamma_A \, S\, \left[\bar{n}\gamma^\mu \gamma_5n + \bar{p}_v\gamma^\mu \gamma_5p_v + \bar{\Lambda}^0_v\gamma^\mu \gamma_5\Lambda^0_v  + \bar{\Xi}^0_v\gamma^\mu \gamma_5 \Xi^0_v +  \bar{\Xi}^-_v\gamma^\mu \gamma_5 \Xi^-_v \right. \nonumber\\
& \qquad \qquad \quad \left.+ \bar{\Sigma}^0_v\gamma^\mu \gamma_5 \Sigma^0_v + \bar{\Sigma}^+_v\gamma^\mu \gamma_5 \Sigma^+_v +  \bar{\Sigma}^-_v\gamma^\mu \gamma_5 \Sigma^-_v  \right] + 
\nonumber\\
&\quad \frac{\partial_\mu \phi}{6\Lambda}  \frac{m_{\pi^\pm}^2}{m_{\pi^\pm}^2-m_\phi^2} \Delta_A \left[3\,(D+F)\,(\bar{p}_v \gamma^\mu \gamma_5 p_v -\bar{n}_v \gamma^\mu \gamma_5n_v) + 3 \,(F-D)\, (\bar{\Xi}_v^0 \gamma^\mu \gamma_5\Xi_v^0  -  \bar{\Xi}_v^-\gamma^\mu \gamma_5 \Xi_v^-)\right.
\nonumber\\
& \left. \qquad \qquad \qquad \qquad \qquad  + 2 \sqrt{3}\, D \,(\bar{\Lambda}_v \gamma^\mu \gamma_5\Sigma^0_v + \bar{\Sigma}_0 \gamma^\mu \gamma_5\Lambda_v^0) + 6\, F\, (\bar{\Sigma}_v^+ \gamma^\mu \gamma_5\Sigma_v^+ - \bar{\Sigma}_v^-\gamma^\mu \gamma_5 \Sigma_v^- ) \right] + 
\nonumber\\
& \quad \frac{\partial_\mu \phi}{6\Lambda} \, \Omega_A \left[(3F-D)\, (\bar{p}_v\gamma^\mu \gamma_5 p_v + \bar{n}_v\gamma^\mu \gamma_5 n_v) - (D+ 3F) \, (\bar{\Xi}^0_v\gamma^\mu \gamma_5 \Xi_v^0 + \bar{\Xi}_v^-\gamma^\mu \gamma_5 \Xi_v^-) \right.
\nonumber\\
& \qquad \qquad \quad\left.+ 2\, D\, (\bar{\Sigma}_v^- \gamma^\mu \gamma_5\Sigma_v^- + \bar{\Sigma}_v^+ \gamma^\mu \gamma_5\Sigma_v^+ + \bar{\Sigma}^0_v \gamma^\mu \gamma_5\Sigma^0_v-  \bar{\Lambda}^0_v \gamma^\mu \gamma_5\Lambda^0_v) )\right] + \\
& \quad \frac{\partial_\mu \phi }{6 \Lambda} \left\{\Xi_A \left[6(D-F)\,\bar{p}_v \gamma^\mu \gamma_5 \Sigma_v^+ + 6\,(D+F)\, \bar{\Sigma}_v^-\gamma^\mu \gamma_5 \Xi_v^-  - \sqrt{6} \,(D+3F)\, \bar{n}_v \gamma^\mu \gamma_5 \Lambda_v^0 \right.\right.
\nonumber\\
& \qquad \qquad \quad \left.\left.- \sqrt{6} \, (D-3F)\, \bar{\Lambda}_v\gamma^\mu \gamma_5 \Xi_v^0- 3 \sqrt{2} \, (D-F) \bar{n}_v\gamma^\mu \gamma_5 \Sigma^0_v - 3\sqrt{2}\, (D+F)\, \bar{\Sigma}_v^0 \gamma^\mu \gamma_5\Xi_v^0 \right] + \text{h.c.} \right\} - 
\nonumber\\
& \quad \frac{\partial_\mu \phi}{2 \Lambda} (q_V^u-q_V^d) \left[\bar{p}_v\gamma^\mu p_v - \bar{n}_v\gamma^\mu n_v  + \bar{\Xi}^0_v\gamma^\mu \Xi^0_v - \bar{\Xi}^-_v\gamma^\mu \Xi^-_v + 2 \bar{\Sigma}^+_v\gamma^\mu \Sigma^+_v - 2\bar{\Sigma}^-_v\gamma^\mu \Sigma^-_v  \right] - 
\nonumber\\
& \quad \frac{\partial_\mu \phi}{2 \Lambda} (q_V^u+q_V^d-2q_V^s) \left[\bar{p}_v\gamma^\mu p_v + \bar{n}_v\gamma^\mu n_v  - \bar{\Xi}^0_v\gamma^\mu \Xi^0_v - \bar{\Xi}^-_v\gamma^\mu \Xi^-_v\right] + 
\nonumber\\
& \quad \frac{\partial_\mu \phi }{2 \Lambda} \left\{Y_V^{ds} \left[\sqrt{6} \bar{\Lambda}_v \gamma^\mu \Xi^0_v -\sqrt{6} \bar{n}_v \gamma^\mu \Lambda_v^0 + \sqrt{2} \bar{n}_v \gamma^\mu \Sigma^0_v - \sqrt{2}\bar{\Sigma}^0_v \gamma^\mu \Xi^0_v + 2 \bar{\Xi}^-_v \gamma^\mu \Sigma^-_v - 2 \bar{p}_v \gamma^\mu \Sigma^+_v\right] + \text{h.c.} \right\} 
\nonumber\, , 
\end{align}
where $F = 0.441(26)$, $D= 0.813(43)$ and $S = 0.405 (62)$ (see \cite{Vonk:2021sit} based on the lattice QCD results in \cite{FlavourLatticeAveragingGroup:2019iem}).

This expression can be further simplified by appropriately choosing the entries of the vectorial matrix $Q_V$; setting $q_V^u = q_V^d = q_V^s = 0$ eliminates completely the last but least line and the previous one from the Lagrangian.

In addition to the contributions in Eq.~\eqref{eq:BarDerALP_1} we should also take into account those stemming from the mixing of neutral mesons into an ALP and stemming from the baryon-meson Lagrangian piece 
\begin{align}
    \mathcal{L}_{B\pi} &\supset  -\frac{D}{2f_\pi} \, \text{Tr} (\bar{B}_v \gamma^\mu \gamma_5 \left\{\partial_\mu \pi, B_v \right\}) -  \frac{F}{2f_\pi} \, \text{Tr} (\bar{B}_v \gamma^\mu \gamma_5 \left[\partial_\mu \pi, B_v \right])\nonumber\\
    &\supset -\frac{\partial_\mu \pi_0}{6f_\pi} \left[D\left(3 \bar{p}_v\gamma^\mu \gamma_5p_v-3 \bar{n}_v\gamma^\mu \gamma_5n_v + 3 \bar{\Xi}^-_v\gamma^\mu \gamma_5 \Xi^-_v - 3 \bar{\Xi}^0_{v}\gamma^\mu \gamma_5\Xi^0_{v} + 2 \sqrt{3}( \bar{\Lambda}^0_v\gamma^\mu \gamma_5\Sigma^0_v +  \bar{\Sigma}^0_v\gamma^\mu \gamma_5\Lambda^0_v)\right)\right.\nonumber\\
    & \qquad \qquad +\left. 3F \left(\bar{p}_v\gamma^\mu \gamma_5p_v - \bar{n}_v\gamma^\mu \gamma_5n_v + 2 \bar{\Xi}^0_v\gamma^\mu \gamma_5\Xi^0_v - \bar{\Xi}^-_v\gamma^\mu \gamma_5\Xi^-_v + 2 \bar{\Sigma}^+_v\gamma^\mu \gamma_5\Sigma^+_v- \bar{\Sigma}^-_v\gamma^\mu \gamma_5\Sigma^-_v\right)\right] \\
    &\quad -\frac{\partial_\mu \eta}{2 \sqrt{3}f_\pi} \left[D\left(\bar{p}_v\gamma^\mu \gamma_5p_v+ \bar{n}_v\gamma^\mu \gamma_5n_v + \bar{\Xi}^-_v\gamma^\mu \gamma_5 \Xi^-_v + \bar{\Xi}^0_{v}\gamma^\mu \gamma_5\Xi^0_{v} + 2  \bar{\Lambda}^0_v\gamma^\mu \gamma_5\Lambda^0_v - 2  \bar{\Sigma}^0_v\gamma^\mu \gamma_5\Sigma^0_v \right.\right.\nonumber\\
    & \qquad \qquad \left. - 2  \bar{\Sigma}^+_v\gamma^\mu \gamma_5\Sigma^+_v - 2  \bar{\Sigma}^-_v\gamma^\mu \gamma_5\Sigma^-_v\right) + \left. 3F \left(-\bar{p}_v\gamma^\mu \gamma_5p_v - \bar{n}_v\gamma^\mu \gamma_5n_v + \bar{\Xi}^0_v\gamma^\mu \gamma_5\Xi^0_v + \bar{\Xi}^-_v\gamma^\mu \gamma_5\Xi^-_v \right)\right] 
    \nonumber\, ,
\end{align}
originating from the first-order expansion of the second and third terms in Eq.~\eqref{eq:Heavy_Baryon_Lagrangian}.
We neglected terms involving neutral kaons as we can choose the entries of the matrix $Q_A$ in such a way to get rid of them. This is, however, not simultaneously possible for the $\pi_0$ and the $\eta$ and one will have the following ALP interaction: 
\begin{align}
    \delta\mathcal{L}_{B\pi}
    &\supset -\xi A\frac{m_\phi^2 + (\alpha'/A)}{m_{\pi_0}^2 - m_\phi^2} \frac{\partial_\mu \phi}{6f_\pi}\cdot \nonumber\\
    & \qquad \cdot\left[D\left(3 \bar{p}_v\gamma^\mu \gamma_5p_v-3 \bar{n}_v\gamma^\mu \gamma_5n_v + 3 \bar{\Xi}^-_v\gamma^\mu \gamma_5 \Xi^-_v - 3 \bar{\Xi}^0_{v}\gamma^\mu \gamma_5\Xi^0_{v} + 2 \sqrt{3}( \bar{\Lambda}^0_v\gamma^\mu \gamma_5\Sigma^0_v +  \bar{\Sigma}^0_v\gamma^\mu \gamma_5\Lambda^0_v)\right)\right. \nonumber\\
    & \qquad \qquad +\left. 3F \left(\bar{p}_v\gamma^\mu \gamma_5p_v - \bar{n}_v\gamma^\mu \gamma_5n_v +  \bar{\Xi}^0_v\gamma^\mu \gamma_5\Xi^0_v - \bar{\Xi}^-_v\gamma^\mu \gamma_5\Xi^-_v + 2 \bar{\Sigma}^+_v\gamma^\mu \gamma_5\Sigma^+_v- 2\bar{\Sigma}^-_v\gamma^\mu \gamma_5\Sigma^-_v\right)\right] \nonumber\\
    &\quad -\xi B\frac{m_\phi^2 + (\beta'/B)}{m_{\eta}^2 - m_\phi^2} \frac{\partial_\mu \phi}{2 \sqrt{3}f_\pi} \cdot \\
    & \qquad \cdot \left[D\left(\bar{p}_v\gamma^\mu \gamma_5p_v+ \bar{n}_v\gamma^\mu \gamma_5n_v + \bar{\Xi}^-_v\gamma^\mu \gamma_5 \Xi^-_v + \bar{\Xi}^0_{v}\gamma^\mu \gamma_5\Xi^0_{v} + 2  \bar{\Lambda}^0_v\gamma^\mu \gamma_5\Lambda^0_v - 2  \bar{\Sigma}^0_v\gamma^\mu \gamma_5\Sigma^0_v \right.\right. \nonumber\\
    & \qquad \qquad \left. - 2  \bar{\Sigma}^+_v\gamma^\mu \gamma_5\Sigma^+_v - 2  \bar{\Sigma}^-_v\gamma^\mu \gamma_5\Sigma^-_v\right) + \left. 3F \left(-\bar{p}_v\gamma^\mu \gamma_5p_v - \bar{n}_v\gamma^\mu \gamma_5n_v + \bar{\Xi}^0_v\gamma^\mu \gamma_5\Xi^0_v + \bar{\Xi}^-_v\gamma^\mu \gamma_5\Xi^-_v \right)\right] 
    \nonumber\, .
\end{align}
With our choice in Eq.~\eqref{eq:Minimal_Mixing_Choice} this amounts to
\begin{align}
    \delta\mathcal{L}_{B\pi} &=  \frac{M_\phi^2}{m_{\eta}^2 - M_\phi^2} 
    \bigg[1 + \frac{m_{\pi^\pm}^2}{m_{\pi^\pm}^2-M_\phi^2}\frac{m_d-m_u}{m_d+m_u}\frac{\Delta_A}{\Omega_A} \bigg] \Omega_A \frac{\partial_\mu \phi}{6\Lambda} \cdot 
    [(3F-D)\, (\bar{p}_v\gamma^\mu \gamma_5 p_v + \bar{n}_v\gamma^\mu \gamma_5 n_v) 
    \nonumber\\
    & -(D+ 3F) \, (\bar{\Xi}^0_v\gamma^\mu \gamma_5 \Xi_v^0 + \bar{\Xi}_v^-\gamma^\mu \gamma_5 \Xi_v^-)
    + 2D (\bar{\Sigma}_v^- \gamma^\mu \gamma_5\Sigma_v^- + \bar{\Sigma}_v^+ \gamma^\mu \gamma_5\Sigma_v^+ + \bar{\Sigma}^0_v \gamma^\mu \gamma_5\Sigma^0_v-  \bar{\Lambda}^0_v \gamma^\mu \gamma_5\Lambda^0_v))] .
\end{align}

Summing this piece with the ones in Eq.~\eqref{eq:BarDerALP_v1} we finally obtain
\begin{align}
\label{eq:BarDerALP_v2}
\mathcal{L}_{\text{ALP}}^{\text{BD, OC}} &= \frac{\partial_\mu \phi}{3\Lambda} \Gamma_A  \, S\,\left[\bar{n}\gamma^\mu \gamma_5n + \bar{p}_v\gamma^\mu \gamma_5p_v + \bar{\Lambda}^0_v\gamma^\mu \gamma_5\Lambda^0_v  + \bar{\Xi}^0_v\gamma^\mu \gamma_5 \Xi^0_v +  \bar{\Xi}^-_v\gamma^\mu \gamma_5 \Xi^-_v \right. \nonumber\\
& \qquad \qquad \quad \left.+ \bar{\Sigma}^0_v\gamma^\mu \gamma_5 \Sigma^0_v + \bar{\Sigma}^+_v\gamma^\mu \gamma_5 \Sigma^+_v +  \bar{\Sigma}^-_v\gamma^\mu \gamma_5 \Sigma^-_v  \right] +
\nonumber\\ 
& \quad \frac{\partial_\mu \phi}{6\Lambda}  \frac{m_{\pi^\pm}^2}{m_{\pi^\pm}^2-m_\phi^2} \Delta_A \left[3\,(D+F)\,(\bar{p}_v \gamma^\mu \gamma_5 p_v -\bar{n}_v \gamma^\mu \gamma_5n_v) + 3 \,(F-D)\, (\bar{\Xi}_v^0 \gamma^\mu \gamma_5\Xi_v^0  -  \bar{\Xi}_v^-\gamma^\mu \gamma_5 \Xi_v^-)\right. 
\nonumber\\
& \left. \qquad \qquad \qquad \qquad \qquad  + 2 \sqrt{3}\, D \,(\bar{\Lambda}_v \gamma^\mu \gamma_5\Sigma^0_v + \bar{\Sigma}_0 \gamma^\mu \gamma_5\Lambda_v^0) + 6\, F\, (\bar{\Sigma}_v^+ \gamma^\mu \gamma_5\Sigma_v^+ - \bar{\Sigma}_v^-\gamma^\mu \gamma_5 \Sigma_v^- ) \right] + 
\nonumber\\
& \quad \frac{\partial_\mu \phi}{6\Lambda} \, \Omega_A\left[\frac{m_\eta^2}{m_\eta^2 - m_\phi^2} + \frac{m_{\pi_{\pm}}^2}{m_{\pi_{\pm}}^2- m_\phi^2} \frac{m_\phi^2}{m_\eta^2-m_\phi^2}\frac{m_d-m_u}{m_d+m_u}\frac{\Delta_A}{\Omega_A}\right]\cdot\\
& \qquad \qquad \qquad \cdot\left[(3F-D)\, (\bar{p}_v\gamma^\mu \gamma_5 p_v + \bar{n}_v\gamma^\mu \gamma_5 n_v) - (D+ 3F) \, (\bar{\Xi}^0_v\gamma^\mu \gamma_5 \Xi_v^0 + \bar{\Xi}_v^-\gamma^\mu \gamma_5 \Xi_v^-) \right.
\nonumber\\
& \qquad \qquad \qquad \quad \left.+ 2\, D\, (\bar{\Sigma}_v^- \gamma^\mu \gamma_5\Sigma_v^- + \bar{\Sigma}_v^+ \gamma^\mu \gamma_5\Sigma_v^+ + \bar{\Sigma}^0_v \gamma^\mu \gamma_5\Sigma^0_v-  \bar{\Lambda}^0_v \gamma^\mu \gamma_5\Lambda^0_v) )\right] + 
\nonumber\\
& \quad \frac{\partial_\mu \phi }{6 \Lambda} \left\{\Xi_A \left[6(D-F)\,\bar{p}_v \gamma^\mu \gamma_5 \Sigma_v^+ + 6\,(D+F)\, \bar{\Sigma}_v^-\gamma^\mu \gamma_5 \Xi_v^-  - \sqrt{6} \,(D+3F)\, \bar{n}_v \gamma^\mu \gamma_5 \Lambda_v^0 \right.\right.
\nonumber\\
& \qquad \qquad \quad \left.\left.- \sqrt{6} \, (D-3F)\, \bar{\Lambda}_v\gamma^\mu \gamma_5 \Xi_v^0- 3 \sqrt{2} \, (D-F) \bar{n}_v\gamma^\mu \gamma_5 \Sigma^0_v - 3\sqrt{2}\, (D+F)\, \bar{\Sigma}_v^0 \gamma^\mu \gamma_5\Xi_v^0 \right] + \text{h.c.} \right\} + 
\nonumber\\
& \quad \frac{\partial_\mu \phi }{2 \Lambda} \left\{Y_V^{ds} \left[\sqrt{6} \bar{\Lambda}_v \gamma^\mu \Xi^0_v -\sqrt{6} \bar{n}_v \gamma^\mu \Lambda_v^0 + \sqrt{2} \bar{n}_v \gamma^\mu \Sigma^0_v - \sqrt{2}\bar{\Sigma}^0_v \gamma^\mu \Xi^0_v + 2 \bar{\Xi}^-_v \gamma^\mu \Sigma^- - 2 \bar{p}_v \gamma^\mu \Sigma^+_v\right] + \text{h.c.} \right\} \, , 
\nonumber
\end{align}
where we set $q_V^u=q_V^d = q_V^s = 0$.

There is a second kind of contribution to the ALP-baryon couplings, stemming from the non-derivative couplings to the quark and gluonic scalar densities. It can be generically parametrized as follows
\begin{equation}
\label{eq:ALP_Baryon_Interactions}
\mathcal{L}_{\text{ALP}}^{\text{Bar., ND}} = \frac{\phi}{\Lambda} C_{\phi \text{BB}} \bar{B}_v B_v \, .
\end{equation}
Since, once again, the ALP has no impact on the hadronic matrix element, i.e.~$\bra{B} \phi \, \theta^\mu_\mu \ket{B} =  \phi \bra{B}\theta^\mu_\mu \ket{B}$, we can then simply compute the effective coupling between an ALP and baryons by requiring the equivalence 

\begin{equation}
    \sum_q \zeta_q \bar{q} q \,\phi +  g_{\phi gg}\, \phi \, GG  \qquad \longrightarrow \qquad   \bra{B}\sum_q \zeta_q \bar{q} q + g_{\phi gg} \, GG \ket{B}\,   \phi \,  \bar{B} B \, , 
\end{equation}
and computing the hadronic matrix element of the scalar quark and gluon densities.

These can be related to the hadronic matrix elements of the proton by making use of the $SU(3)_V$ symmetry underlying the construction of the baryonic Lagrangian itself \cite{Gasser:1982ap}. In particular, it emerges that the matrix element of the scalar gluonic density is approximately universal, i.e.
\begin{equation}
    \bra{B} GG \ket{B} = \bra{p} GG \ket{p} = -\frac{8\pi}{9 \alpha_s} (m_p - \sigma_u - \sigma_d - \sigma_s) \, , \quad \text{with} \quad B= \{p, n, \Sigma^+, \Sigma^-, \Sigma^0, \Xi^-, \Xi^0, \Lambda^0\} \, , 
\end{equation}
 while one obtains 
 \begin{equation}
 \begin{split}
 &\bra{p} \bar{u} u \ket{p} = \frac{\sigma_u}{m_u} \, , \qquad  \bra{p} \bar{d} d \ket{p} = \frac{\sigma_d}{m_d} \, , \qquad  \bra{p} \bar{s} s \ket{p} = \frac{\sigma_s}{m_s} \, , \\
 &\bra{n} \bar{u} u \ket{n} = \frac{\sigma_d}{m_d} \, ,  \qquad  \bra{n} \bar{d} d \ket{n} = \frac{\sigma_u}{m_u} \, , \qquad  \bra{n} \bar{s} s \ket{n} = \frac{\sigma_s}{m_s} \, , \\
 \end{split}
 \end{equation}
for the nucleonic doublet, as it was the case in a two-flavour setting. As far as the remaining baryons are concerned one instead has 
 \begin{equation}
 \begin{split}
  &\bra{\Sigma^0} \bar{u} u \ket{\Sigma^0} = \frac{1}{2}\left(\frac{\sigma_u}{m_u}+\frac{\sigma_s}{m_s}\right) \qquad  \bra{\Sigma^0} \bar{d} d \ket{\Sigma^0} = \frac{1}{2}\left(\frac{\sigma_u}{m_u}+\frac{\sigma_s}{m_s}\right) \qquad  \bra{\Sigma^0} \bar{s} s \ket{\Sigma^0} = \frac{\sigma_d}{m_d}\\
 &\bra{\Sigma^+} \bar{u} u \ket{\Sigma^+} = \frac{\sigma_u}{m_u} \qquad  \bra{\Sigma^+} \bar{d} d \ket{\Sigma^+} = \frac{\sigma_s}{m_s} \qquad  \bra{\Sigma^+} \bar{s} s \ket{\Sigma^+} = \frac{\sigma_d}{m_d}\\
 &\bra{\Sigma^-} \bar{u} u \ket{\Sigma^-} = \frac{\sigma_s}{m_s} \qquad  \bra{\Sigma^-} \bar{d} d \ket{\Sigma^-} = \frac{\sigma_u}{m_u} \qquad  \bra{\Sigma^-} \bar{s} s \ket{\Sigma^-} = \frac{\sigma_d}{m_d}
 \end{split}
 \end{equation}

 \begin{equation}
 \begin{split}
 &\bra{\Xi^0} \bar{u} u \ket{\Xi^0} = \frac{\sigma_d}{m_d} \quad \quad  \, \, \, \bra{\Xi^0} \bar{d} d \ket{\Xi^0} = \frac{\sigma_s}{m_s} \qquad \! \quad   \bra{\Xi^0} \bar{s} s \ket{\Xi^0} = \frac{\sigma_u}{m_u}\\
 &\bra{\Xi^-} \bar{u} u \ket{\Xi^-} = \frac{\sigma_s}{m_s} \qquad \,  \bra{\Xi^-} \bar{d} d \ket{\Xi^-} = \frac{\sigma_d}{m_d} \qquad \,  \bra{\Xi^-} \bar{s} s \ket{\Xi^-} = \frac{\sigma_u}{m_u}\\
 &\bra{\Lambda^0} \bar{u} u \ket{\Lambda^0} = \frac{1}{6}\left(\frac{\sigma_u}{m_u}+\frac{\sigma_s}{m_s}+\frac{4\sigma_d}{m_d}\right) \quad  \bra{\Lambda^0} \bar{d} d \ket{\Lambda^0} =  \frac{1}{6}\left(\frac{\sigma_u}{m_u}+\frac{\sigma_s}{m_s}+\frac{4\sigma_d}{m_d}\right) \\
 &  \bra{\Lambda^0} \bar{s} s \ket{\Lambda^0} =  \frac{1}{3}\left(\frac{2\sigma_u}{m_u}+\frac{2\sigma_s}{m_s}-\frac{\sigma_d}{m_d}\right) \, . \\
 \end{split}
 \end{equation}
These results can be then directly applied to Eq.~\eqref{eq:AAA} to extract the following ALP-Baryon scalar couplings:
\begin{equation}
\begin{split}
\label{eq:cpp3f}
&C_{\phi \text{pp}} =  \frac{v}{m_u} y_{q,S}^u \, \sigma_u +  \frac{v}{m_d} y_{q,S}^d \, \sigma_d +  \frac{v}{m_s} y_{q,S}^s\,  \sigma_s - \frac{32\pi^2 C_g}{9} \,  (m_p - \sigma_u -\sigma_d-\sigma_s) \, ,  \\
&C_{\phi \text{nn}} =  \frac{v}{m_d} y_{q,S}^u \, \sigma_d +  \frac{v}{m_u} y_{q,S}^d \, \sigma_u +  \frac{v}{m_s} y_{q,S}^s\,  \sigma_s - \frac{32\pi^2 C_g}{9} \,  (m_p - \sigma_u -\sigma_d-\sigma_s) \, , 
\\
&C_{\phi\Sigma^+\Sigma^+} =  \frac{v}{m_u} y_{q,S}^u \, \sigma_u +  \frac{v}{m_s} y_{q,S}^d \, \sigma_s +  \frac{v}{m_d} y_{q,S}^s\,  \sigma_d - \frac{32\pi^2 C_g}{9} \,  (m_p - \sigma_u -\sigma_d-\sigma_s) \, ,  \\
&C_{\phi\Sigma^0\Sigma^0} =  \frac{v}{2} y_{q,S}^u \, \left(\frac{\sigma_u}{m_u}+\frac{\sigma_s}{m_s}\right) +  \frac{v}{2} y_{q,S}^d \, \left(\frac{\sigma_u}{m_u}+\frac{\sigma_s}{m_s}\right) +  \frac{v}{m_d} y_{q,S}^s\,  \sigma_d - \frac{32\pi^2 C_g}{9} \,  (m_p - \sigma_u -\sigma_d-\sigma_s) \, ,  \\
&C_{\phi\Sigma^-\Sigma^-} =  \frac{v}{m_s} y_{q,S}^u \, \sigma_s +  \frac{v}{m_u} y_{q,S}^d \, \sigma_u +  \frac{v}{m_d} y_{q,S}^s\,  \sigma_d - \frac{32\pi^2 C_g}{9} \,  (m_p - \sigma_u -\sigma_d-\sigma_s) \, ,
\end{split}
\end{equation}
\begin{equation}
\begin{split}
\label{eq:cpp3f_bis}
&C_{\phi\Xi^0\Xi^0} =  \frac{v}{m_d} y_{q,S}^u \, \sigma_d +  \frac{v}{m_s} y_{q,S}^d \, \sigma_s +  \frac{v}{m_u} y_{q,S}^s\,  \sigma_u - \frac{32\pi^2 C_g}{9} \,  (m_p - \sigma_u -\sigma_d-\sigma_s) \, ,  \\
&C_{\phi\Xi^-\Xi^-} =  \frac{v}{m_s} y_{q,S}^u \, \sigma_s +  \frac{v}{m_d} y_{q,S}^d \, \sigma_d +  \frac{v}{m_u} y_{q,S}^s\,  \sigma_u - \frac{32\pi^2 C_g}{9} \,  (m_p - \sigma_u -\sigma_d-\sigma_s) \, ,  \\
&C_{\phi\Lambda^0\Lambda^0} =  \frac{v}{6} y_{q,S}^u \, \left(\frac{\sigma_u}{m_u}+\frac{\sigma_s}{m_s}+ 4\frac{\sigma_d}{m_d}\right) +  \frac{v}{6} y_{q,S}^d \, \left(\frac{\sigma_u}{m_u}+\frac{\sigma_s}{m_s}+ 4\frac{\sigma_d}{m_d}\right) +  \frac{v}{3} y_{q,S}^s\,  \left(2\frac{\sigma_u}{m_u}+2\frac{\sigma_s}{m_s}-\frac{\sigma_d}{m_d}\right)\\
& \qquad \qquad - \frac{32\pi^2 C_g}{9} \,  (m_p - \sigma_u -\sigma_d-\sigma_s) \, ,
\end{split}
\end{equation}
where the explicit values for the various quantities and the procedure to obtain them can be found in \cite{Cheng:2012qr}.

\subsection{Summary and Jarlskog invariants}

To summarize, we built the low-energy effective Lagrangian describing ALP interactions with mesons, baryons and photons, finding that it can be decomposed into two pieces with opposite CP transformation properties 
\begin{align}
\label{eq:Simplest_Lag_3f_CPE}
(\mathcal{L}_{\text{ALP}}^{\chi\text{pt}})^{\rm CP-even} &= e^2 \frac{c_\gamma}{\Lambda} \, \phi \,  F^{\mu\nu}F_{\mu\nu} 
+ \mathcal{L}_{\text{ALP}}^{\kappa} + \mathcal{L}_{\text{ALP}}^{\mathcal{Z}} + \mathcal{L}_{\text{ALP}}^{\text{Bar. ND}}
\\
(\mathcal{L}_{\text{ALP}}^{\chi\text{pt}})^{\rm CP-odd} &=  e^2 \frac{\tilde{c}_\gamma}{\Lambda} \, \phi \, F^{\mu\nu}\tilde{F}_{\mu\nu} +\mathcal{L}_{\text{ALP}}^{\text{MI}}+\mathcal{L}_{\text{ALP}}^{\text{ADI}} + \mathcal{L}_{\text{ALP}}^{\eta \text{KT}+\text{MT}} + \mathcal{L}_{\text{ALP}}^{\text{BD, OC}} \, , 
\label{eq:Simplest_Lag_3f_CPO}
\end{align}
where the explicit expressions for the terms in the above Lagrangians have been provided previously. 
The associated Jarlskog invariants are reported in Table \ref{tab:3F_Jarlskog}. To simplify their interpretation, we also report below the compositions of all the coefficients appearing in Table \ref{tab:3F_Jarlskog} in terms of the 
microscopic parameters of Eq.~\ref{eq:AAA}
\begin{eqnarray}
    c_\gamma  &=& \, C_\gamma - \frac{\beta^0_{\text{QED}}}{\beta^0_{\text{QCD}}} C_g \, ,\\
    \tilde{c}_\gamma &=& \, \tilde{C}_\gamma  -4 N_c \text{ tr }(Q_AQ_q^2)  \tilde{C}_g  \, , \\
    \kappa &=& \frac{8\pi}{\alpha_s}\frac{g_s^2 C_g}{\beta^0_\text{QCD}} \, ,\\
    \mathcal{Z} &=& y_{S}  + \frac{g_s^2 C_g }{\beta^0_{\text{QCD}}} \frac{8\pi}{\alpha_s} \frac{M_q}{v} \, , \\
    \Delta_A &=& Y_P^u -Y_P^d - 16 \pi^2 \tilde{C}_g\frac{m_s(m_d-m_u)}{m_u m_d + m_u m_s + m_d m_s}  \, , \\
    \Omega_A &=& Y_P^u + Y_P^d -2Y_P^s - \frac{\lambda_g}{2} \frac{m_s(m_d+m_u)- 2 m_u m_d}{m_m m_d + m_u m_s + m_d m_s} \, , \\
    \label{eq:Xidef}
    \Xi_A &=&  Y_{P}^{ds}\frac{m_{\pi^\pm}^2 (m_d+m_s)}{(m_d+m_s) m_{\pi^{\pm}}^2 - m_\phi^2(m_u+m_d)} \, , \\
    \Gamma_A &=& Y_P^u + Y_P^d + Y_P^s - \frac{\lambda_g}{2} \, ,\\
    \label{eq:YVdef}
    Y_V^{ds} &=& Y_S^{ds}\, ,\\
    C_{\phi \text{BB}} &=& \text{see equations~\eqref{eq:cpp3f} 
    and ~\eqref{eq:cpp3f_bis}}\, .
\end{eqnarray}
\begin{table}[ht!]
    \centering
    \begin{tabular}{c|c|c|c|c|c|c}
                            & $c_\gamma$ & $y_{\ell,S}$&  $\kappa$& $\mathcal{Z}$& $Y_V^{ds}$& $C_{\phi \text{BB}}$ \\
                            \hline
      $\tilde{c}_\gamma$    & $ \tilde{c}_\gamma\,  c_\gamma $ & $ \tilde{c}_\gamma\,  y_{\ell,S}$ & $ \tilde{c}_\gamma\,  \kappa $ & $ \tilde{c}_\gamma\,  \mathcal{Z} $ & $ \tilde{c}_\gamma\,  Y_V^{ds} $ & $ \tilde{c}_\gamma\, C_{\phi \text{BB}} $\\
      $y_{\ell, P}$    & $ y_{\ell, P}\,  c_\gamma $ & $ y_{\ell, P}\,  y_{\ell,S}$ & $ y_{\ell, P}\,  \kappa $ & $ y_{\ell, P}\,  \mathcal{Z} $& $ y_{\ell, P}\,  Y_V^{ds} $ & $ y_{\ell, P}\,  C_{\phi \text{BB}} $\\
      $\Delta_A$    & $ \Delta_A\,  c_\gamma $ & $ \Delta_A\,  y_{\ell,S}$ & $ \Delta_A\,  \kappa $ & $ \Delta_A\,  \mathcal{Z} $ & $ \Delta_A\,  Y_V^{ds} $& $ \Delta_A\, C_{\phi \text{BB}} $\\
      $\Omega_A$    & $ \Omega_A\,  c_\gamma $ & $ \Omega_A\,  y_{\ell,S}$ & $ \Omega_A\,  \kappa $ & $ \Omega_A\,  \mathcal{Z} $ & $ \Omega_A\,  Y_V^{ds} $& $ \Omega_A\, C_{\phi \text{BB}} $\\
      $\Xi_A$    & $ \Xi_A\,  c_\gamma $ & $ \Xi_A\,  y_{\ell,S}$ & $ \Xi_A\,  \kappa $ & $ \Xi_A\,  \mathcal{Z} $ & $ \Xi_A\,  Y_V^{ds} $& $ \Xi_A\, C_{\phi \text{BB}} $\\
      $\Gamma_A$    & $ \Gamma_A\,  c_\gamma $ & $ \Gamma_A\,  y_{\ell,S}$ & $ \Gamma_A\,  \kappa $ & $ \Gamma_A\,  \mathcal{Z} $ & $ \Gamma_A\,  Y_V^{ds} $& $ \Gamma_A\, C_{\phi \text{BB}}$
    \end{tabular}
    \caption{Jarlskog invariants emerging from the interactions in \eqref{eq:Simplest_Lag_3f_CPE} and \eqref{eq:Simplest_Lag_3f_CPO}}
    \label{tab:3F_Jarlskog}
\end{table}

\section{Phenomenological applications}
\label{sec:pheno}

Although a detailed phenomenological analysis is beyond the scope of the present work, in the following we will briefly discuss a few interesting CP-violating observables which can be analysed by means of the ALP interactions derived above in the the $N_f=2$ and $N_f=3$ frameworks.\footnote{For future phenomenological applications, fully customisable \texttt{FeynRules}~\cite{Alloul:2013bka} models 
both for $N_f =2$ and $N_f =3$ settings are provided as ancillary files.}

\subsection{Proton and neutron EDM}

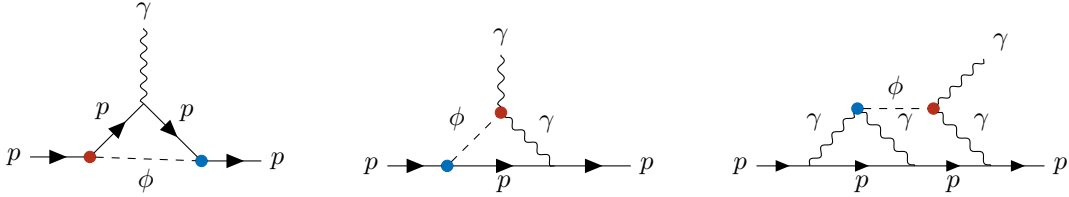
\begin{figure}[h]
\centering
	\begin{tikzpicture}[baseline = (a)]
	\begin{feynman}[]
	\vertex (a){$p$};
    \node [ BrickRed, dot, right = 1  of a](b);
	\vertex [above right = 1 of b] (c) ;
	\node [ NavyBlue, dot,  right= 2.4 of a] (d);
	\vertex [right= 1 of d] (e){$p$};
    \vertex [above= 0.75 of c] (f){$\gamma$};
	\diagram*{
		(a) -- [fermion] (b),
  		(b) -- [fermion, edge label = \(p\)] (c),
      	(c) -- [fermion, edge label = \(p\)] (d),
	  	(d) -- [fermion] (e),
        (b) -- [scalar, edge label' = \(\phi\)] (d),
        (f) -- [photon] (c)	
        };
	\end{feynman}
	\end{tikzpicture}
 \qquad
 	\begin{tikzpicture}[baseline = (a)]
	\begin{feynman}[]
	\vertex (a){$p$};
    \node [NavyBlue, dot, right = 1  of a](b);
	\node [BrickRed, dot, above right = 1 of b] (c) ;
	\vertex [below right= 1 of c] (d);
	\vertex [right= 1 of d] (e){$p$};
    \vertex [above= 1 of c] (f){$\gamma$};
	\diagram*{
		(a) -- [fermion] (b),
  		(b) -- [scalar, edge label = \(\phi\)] (c),
      	(c) -- [photon, edge label = \(\gamma\)] (d),
	  	(d) -- [fermion] (e),
        (b) -- [fermion, edge label' = \(p\)] (d),
        (f) -- [photon] (c)
	};
	\end{feynman}
	\end{tikzpicture}
 \qquad 
 \begin{tikzpicture}[baseline = (a)]
	\begin{feynman}[small]
	\vertex (a){$p$};
    \vertex [right= 0.9 of a](b);
    \vertex [right= 1.4 of b](c);
    \node [NavyBlue, dot,scale = 1.5, above left = of c](e);
	\node [BrickRed, dot,scale = 1.5, right = of e] (f) ;
	\vertex [right= of c] (d);
	\vertex [right= 0.7 of d] (g){$p$};
    \vertex [above right = 1.25 of f] (h){$\gamma$};
	\diagram*{
		(a) -- [fermion] (b),
  	    (b) -- [fermion, edge label' = \(p\)] (c),
  	    (c) -- [fermion, edge label' = \(p\)] (d),
  	    (d) -- [fermion] (g),
      	(b) -- [photon, edge label = \(\gamma\)] (e),
        (c) -- [photon, edge label' = \(\gamma\)] (e),
        (d) -- [photon, edge label' = \(\gamma\)] (f),
      	(f) -- [photon] (h),
        (e) -- [scalar, edge label = \(\phi\)] (f)
	};
	\end{feynman}
	\end{tikzpicture}
\caption{Feynman diagrams contributing to the ALP-mediated effects to the proton EDM, $d_p$.}
\label{fig:pEDMALPDiagrams}
\end{figure}

The effective Lagrangian given in \eqref{eq:Simplest_Lag_2f} immediately allows one to compute the contributions of a 
light ALP to the proton EDM.  
Here, we assume that the contribution to EDMs 
originating from a possible ALP vacuum expectation value 
is absent thanks to a UV mechanism addressing the strong CP problem. Then,
working in the basis without ALP-pion mixing, the  
leading 
topologies are the ones in Fig.~\ref{fig:pEDMALPDiagrams}.

Defining the electric dipole moment of a fermion $\psi$ via $\mathcal{L}_{\text{dip.}} = -i \,(d_\psi/2)\, \bar{\psi} \sigma^{\mu\nu} \gamma_5 \psi F_{\mu\nu}$, we find that
\begin{equation}
d_p \simeq -\frac{e \, Q_p}{4\pi^2 \Lambda^2} 
\left[C_{\phi\text{pp}}\tilde{C}_{\phi \text{p}}  + 
e^2 m_p c_\gamma \tilde{C}_{\phi \text{p}} 
\left( 6 + 2\ln \frac{\Lambda_{\text{ren}}^2}{m_p^2}\right) +   e^2 \tilde{c}_\gamma C_{\phi \text{pp}} 
\left(2 + \ln \frac{\Lambda_{\text{ren}}^2}{m_p^2} \right)
+ 3 \frac{Q_p^2}{\pi^2} m_p e^6 c_\gamma \tilde{c}_\gamma \ln^2 \frac{\Lambda_{\text{ren}}}{m_\phi}
\right] \, ,
\end{equation}
%
%
where we worked in the limit $m_\phi \ll m_p$ and  the renormalization scale is set to $\Lambda_{\rm ren}\simeq m_p$. 

A natural question is whether a similar effect can be induced for the neutron EDM, $d_n$. Actually, no such contribution can be generated at leading order in $\chi$pt, due to the fact that neutrons have null electric charge and they do not have any minimal coupling to the electromagnetic field-strength tensor.
Nonetheless, couplings of neutrons to photons are generated at next-to-leading order in $\chi$pt \cite{Fettes:1998ud,Oller:2006yh}. In the $N_f=2$ case, the 
relevant relativistic Lagrangian reads
\begin{equation}
\mathcal{L}_{\gamma \text{N}}^{\text{NLO}} = 
\frac{F_{\mu\nu}}{4m_N}\left[\, \mathcal{C}_p \,\bar{p}\sigma^{\mu\nu}p + \mathcal{C}_n \,\bar{n}\sigma^{\mu\nu}n \,\right] \, , 
\end{equation}
where $\mathcal{C}_p \simeq 1.79$~\cite{ParticleDataGroup:2022pth} and $\mathcal{C}_n \simeq 1.91$~\cite{ParticleDataGroup:2022pth} are measured low-energy constants.
Given these interactions one can then compute the corresponding contribution to $d_n$, whose Feynman diagrams are the same as for the proton. We find the following order of magnitude estimate: 
\begin{equation}
d_n \sim -\frac{e \, \mathcal{C}_n}{4\pi^2 \Lambda^2} 
\left[ -\frac{3}{8}C_{\phi\text{nn}}\tilde{C}_{\phi \text{n}}
\frac{\Lambda_{\text{ren}}^2}{m_n^2} 
+ e^2 (2 m_n c_\gamma \tilde{C}_{\phi \text{n}}+ 3 \tilde{c}_\gamma C_{\phi \text{nn}}) \ln \frac{\Lambda_{\text{ren}}^2}{m_n^2} \right]\,.
\end{equation}
Note that $d_p$ and $d_n$ have a comparable size while their
current experimental bounds read $d_p < 2.1 \times 10^{-25}\,e$ cm~\cite{Sahoo:2016zvr}
and $d_n < 1.8 \times 10^{-26}\,e$ cm (90\% C.L.)~\cite{Abel:2020pzs,Pendlebury:2015lrz}.
%
By turning on just two couplings at a time and imposing these bounds, we can constrain some of the low-energy Jarlskog invariants of the theory, see Table \ref{tab:constraints}.

%

\begin{table}[ht!]
    \centering
    \begin{tabular}{|c|c|c|c|c|}
    \hline
                           & $y_S^u$ & $y_S^d$ & $C_\gamma$ & $C_g$\\
                           \hline
      $y_P^u$                & $ 2.8 \times 10^{-12} $ & $ 1.8 \times 10^{-12}$   & $ 8.0 \times 10^{-8}$  & $ 1.1 \times 10^{-10}$\\
      \hline
      $y_P^d$                & $ 9.9 \times 10^{-12}$ & $ 6.4 \times 10^{-12} $  & $ 1.1 \times 10^{-7}$  & $ 3.8 \times 10^{-10}$\\
      \hline
      $\tilde{C}_\gamma$     & $ 1.2 \times 10^{-6}$ & $ 1.8 \times 10^{-6}$  & $  4.8\times 10^{-3}$ & $ 7.1 \times 10^{-5}$\\
      \hline
      $\tilde{C}_g$          & $ 1.1 \times 10^{-9}$ & $ 7.4 \times 10^{-10}$ & $ 2.4 \times 10^{-5}$  & $ 4.4 \times 10^{-8}$\\ \hline
    \end{tabular}
    \caption{Upper limits on Jarlskog invariants obtained 
    from the bounds on the neutron and proton EDMs assuming $\Lambda = 1$ TeV. To make contact with the notation in Ref.~\cite{DiLuzio:2020oah} here we have defined $y_{P}^{q} = - 2\frac{m_q}{v} Y_P^q$.}
    \label{tab:constraints}
\end{table}

\subsection{CP-violating ALP decays
}
A significant observable probing CP-violating effects is given by the ratio
\begin{equation}
r (i,j) \equiv \frac{\text{BR}\left(\phi \rightarrow \pi_i \pi_j\right)}{\text{BR} \left(\phi \rightarrow  \pi_i \pi_j \pi_0\right)}\,,\qquad\qquad (i,j) = \{(+-), (00)\}\,, 
\end{equation}
which could be relevant in phenomenological analyses 
of hadronically coupled GeV-scale ALPs, see e.g.~\cite{Aloni:2018vki}.
The associated scattering amplitudes can in turn 
be computed by making use of the Feynman 
rules extracted in a two-flavour setting from the Lagrangian in \eqref{eq:Simplest_Lag_2f}. Starting from 
\begin{equation}
\begin{split}
i \mathcal{M} (\phi(Q)\rightarrow \pi^+(p_1) \pi^-(p_2)) &= i \frac{m_\pi^2}{\Lambda} \left[ v \frac{\mathcal{Z}_d + \mathcal{Z}_u}{m_d+m_u} - \kappa \left(2+ \frac{m_\phi^2}{m_\pi^2}\right) \right] \, , \\
i \mathcal{M} (\phi (Q) \rightarrow \pi^0 (p_1)\pi^0(p_2)) &= i \frac{m_\pi^2}{\Lambda} \left[ v \frac{\mathcal{Z}_d + \mathcal{Z}_u}{m_d+m_u} - \kappa \left(2+ \frac{m_\phi^2}{m_\pi^2}\right) \right] \, , \\
i \mathcal{M} (\phi (Q) \rightarrow \pi^+(p_1) \pi^-(p_2) \pi^0(p_3)) &= -i \frac{\Delta_{ud}^A}{ f_\pi \Lambda} \frac{m_\pi^2}{(m_\phi^2-m_\pi^2)} \left[ m_\phi^2 - 2\, Q \cdot p_3 \right] \, , \\
i \mathcal{M} (\phi (Q) \rightarrow \pi^0 (p_1) \pi^0(p_2)\pi^0(p_3)) &= -i \frac{\Delta_{ud}^A}{ f_\pi \Lambda} \frac{m_\pi^2 m_\phi^2}{(m_\phi^2-m_\pi^2)}  \, , 
\end{split}
\end{equation}
the decay rates of two-body and three-body processes are then computed making use of the formulas: 
\begin{equation}
\begin{split}
\Gamma_{\phi \rightarrow \pi_i \pi_j} &= \frac{g_{ij}}{64\pi^2}\frac{1}{m_\phi} \sqrt{1- \frac{4m_\pi^2}{m_\phi^2}} \int_0^{2\pi} d\varphi \int_{-1}^1 d\cos \theta |\bar{\mathcal{M}}|^2 \, , \\
\Gamma_{\phi \rightarrow \pi_i \pi_j \pi_0} &= \frac{m_\phi}{256\pi^3} g_{ij0}  \int_{x_1^{\text{max}}}^{x_1^{\text{min}}} dx_1  \int_{x_2^{\text{max}}}^{x_2^{\text{min}}} dx_2|\bar{\mathcal{M}}|^2 \, , 
\end{split}
\end{equation}
where $g_{+-} = g_{+-0} = 1$, $g_{00}=1/2$ and $g_{000} = 1/6$ are the symmetry factors for the final-state particles. 
In the three-body final space the variables $x_i$ are defined via $x_i = 2 Q\cdot p_i/m_\phi^2$, while the extrema of integration are
\begin{equation}
x_2^{\text{max,min}} = \frac{x_1^2-3 x_1+2-r_\pi (x_1-2) \pm \sqrt{-3 r_\pi^2+2 r_\pi (x_1-1)+(x_1^2-1)^2} \sqrt{x_1^2-4 r_\pi}}{2 (r_\pi-x_1+1)} \, ,
\end{equation}
and
\begin{equation}
x_1^{\text{max}} = 1- 3 r_\pi \, , \qquad\qquad
x_1^{\text{min}} = 2 \sqrt{r_\pi} \, ,
\end{equation}
where we have defined $r_\pi = m_\pi^2/m_\phi^2$.
As a consequence one finds
\begin{equation}
\begin{split}
&r(0,0) = 48 \pi^2 \frac{f_\pi^2}{m_\phi^2} 
\frac{(1-r_\pi)^2 \sqrt{1-4 r_\pi}}{r_\pi^2} 
\frac{\left|v r_\pi \frac{\mathcal{Z}_u + \mathcal{Z}_d}{m_u + m_d} - \kappa \, \left(1 + 2r_\pi\right)\right|^2}{|\Delta_{ud}^A|^2}\frac{1}{\int dx_1 dx_2} \, , \\
&r(+,-) = 16 \pi^2 \frac{f_\pi^2}{m_\phi^2} 
\frac{(1-r_\pi)^2 \sqrt{1-4 r_\pi}}{r_\pi^2} 
\frac{\left|v r_\pi \frac{\mathcal{Z}_u + \mathcal{Z}_d}{m_u + m_d} - \kappa \, \left(1 + 2r_\pi\right)\right|^2}{|\Delta_{ud}^A|^2}\frac{1}{\int dx_1 dx_2 (x_1+x_2-1)^2} \, , 
\end{split}
\end{equation}
where we have used the fact that $x_1+x_2+x_3 = 2$ by definition.
As an example, we evaluate the above ratios for $m_\phi = m_K$,
obtaining the following numerical estimates 
\begin{equation}
r(0,0) \simeq 2.5 \times 10^4 
\left|\frac{C_g - 82 \, y_S^d- 82 \, y_S^u}{\tilde{C}_g-454 \, y_P^d+ 981 \, y_P^u}\right|^2\,, \qquad 
r(+,-) \simeq 7.4 \times 10^4 \left|\frac{C_g - 82 \, y_S^d- 82 \, y_S^u}{\tilde{C}_g-454 \, y_P^d+ 981 \, y_P^u}\right|^2\, .
\end{equation}

\subsection{Kaon decays}

Kaon decays are observables of paramount interest when dealing with CP violation in the mesonic sector.
In our setup, the relevant Lagrangian describing charged kaon interactions is the following
\begin{equation}
\label{eq:ChargedKaons}
\begin{split}
\mathcal{L}_{\text{ALP, KC}}^{\chi\text{pt}} &=  \left[i \frac{Y_V^{ds}}{2\Lambda} \,  \partial \phi \,  \left( 2 DK^- \pi^+ - 2 D\pi^+ K^- \right)  + \frac{\Xi_A}{f_\pi \Lambda} \partial \phi  \left( \partial \pi_0 K^- \pi^+ - D \pi^+ K^- \pi_0 \right) + \text{h.c.} \right] \, , 
\end{split}
\end{equation}
while in the neutral sector, choosing the basis in \eqref{eq:Minimal_Mixing_Choice} in order to minimize mesonic mixing effects, we have
\begin{align}
\label{eq:NeutralKaons}
\mathcal{L}_{\text{ALP, KN}}^{\chi\text{pt}} &=  
\frac{\partial \phi}{6 f_\pi \Lambda} \left[\Xi_A \left( 4 \sqrt{2} D\pi^- \pi^+ \bar{K}_0 - 2\sqrt{2} D\pi^+ \pi^- \bar{K}_0 +\sqrt{2} \partial  \pi_0  \bar{K}_0 \pi_0   +\sqrt{2}\,  \partial \bar{K}_0 (-2 \pi^+ \pi^-  - \pi_0^2)\right) + \text{h.c.} \right] 
\nonumber\\
& + 
\frac{M_\phi^2}{ f_\pi \Lambda} \, \phi \, \left( \Xi_A\bar{K_0} + \Xi_A^*K_0\right) \left(  2 \pi^+\pi^- + \pi_0^2 \right) + \left[i \frac{Y_V^{ds}}{2\Lambda} \,  \partial \phi \, \left( + \sqrt{2} \partial \pi_0 \bar{K}_0 - \sqrt{2} \partial \bar{K}_0 \pi_0 \right) +\text{h.c.}\right] \, , 
\end{align}
Passing to the mass basis, \eqref{eq:NeutralKaons} becomes
\begin{align}
\label{eq:NeutralKaonsMassBasisSimp}
\mathcal{L}_{\text{ALP, KN}}^{\chi\text{pt}} &=  -2\sqrt{2}\frac{M_\phi^2}{ f_\pi \Lambda} \, \phi \, \left(  2 \pi^+\pi^- + \pi_0^2 \right) \, \left[ \text{ Re } \Xi_A \, K_L + i \text{ Im } \Xi_A \,K_S \right]+\frac{\partial \phi}{3 f_\pi \Lambda}  \, \partial  \pi_0  \pi_0 \left[ \text{ Re } \Xi_A \, K_L + i \text{ Im } \Xi_A\, K_S \right]  \nonumber\\
& \quad -\frac{1}{\Lambda} \,  \partial \phi \, \partial \pi_0  \left[i \text{ Im } Y_V^{ds} \,K_L +\text{ Re } Y_V^{ds} \,  K_S  \right] +\frac{1}{\Lambda} \,  \partial \phi \,  \pi_0  \left[i \text{ Im } Y_V^{ds} \, \partial K_L  +\text{ Re } Y_V^{ds}\, \partial K_S  \right] \nonumber\\
& \quad + \frac{ \sqrt{2} }{3 f_\pi \Lambda} \partial \phi \left[\Xi_A (2 D\pi^- \pi^+ - D\pi^+ \pi^- ) (K_L-K_S) + \Xi_A^*  (2 D\pi^+ \pi^- - D\pi^- \pi^+)  (K_S + K_L) \right] \,,
\end{align}
where the parameters 
$Y_V^{ds}$ and $\Xi_A$ are defined in terms of 
$Y^{ds}_S$ and $Y^{ds}_{P}$ via Eq.~(\ref{eq:YVdef}) 
and (\ref{eq:Xidef}), respectively.

We can then extract the decay rates for the two-body final state decays of either charged or neutral kaons: 
\begin{equation}
\begin{split}
\Gamma (K^+ \to \pi^+ \phi) &= \frac{|Y_V^{ds}|^2}{\Lambda^2} \frac{m_K^3}{16\pi} \left(1-\frac{m_\pi^2}{m_K^2}\right)^2 \, \sqrt{1-2\frac{(m_\pi^2+m_\phi^2)}{m_K^2}+ \frac{(m_\phi^2-m_\pi^2)^2}{m_K^4}} \, , \\
\Gamma (K_L \to \pi_0 \, \phi) &= \frac{|\text{Im }Y_V^{ds}|^2}{\Lambda^2} \frac{m_K^3}{16\pi} \left(1-\frac{m_\pi^2}{m_K^2}\right)^2 \, \sqrt{1-2\frac{(m_\pi^2+m_\phi^2)}{m_K^2}+ \frac{(m_\phi^2-m_\pi^2)^2}{m_K^4}} + \mathcal{O}(\varepsilon) \, .
\end{split}
\end{equation}
Assuming $m_\phi \ll m_K$, we can then employ the BNL result, $\text{BR}(K^+ \to \pi^+ \phi) < 7.3 \times 10^{-11}$ \cite{E949:2007xyy} and the KOTO one, $\text{BR}(K_L \to \pi_0 \phi) < 2 \times 10^{-9}$ \cite{KOTO:2018dsc}, to put bounds on the coupling $Y_V^{ds}$ appearing in the previous expressions: 
\begin{equation}
\begin{split}
|Y_V^{ds}| \lesssim 1.4 \times 10^{-9} \frac{\Lambda}{\text{TeV}}\qquad \text{ and } \qquad 
|\text{Im }Y_V^{ds}| \lesssim 3.6 \times 10^{-9} \frac{\Lambda}{\text{TeV}}\,.
\end{split}
\end{equation}
As far as future perspectives are concerned, the NA62 experiment is expected to improve by one order of magnitude the bounds on $\text{BR}(K^+ \to \pi^+ \phi)$,  
while the KOTO 
experiment foresees a possible improvement to 
$\text{BR}(K_L \to \pi_0 \phi) \lesssim  10^{-11}$, 
see e.g.~Ref.~\cite{MartinCamalich:2020dfe}.

Similarly, one can compute the three-body decay rates of both charged and neutral kaons. We find that 
\begin{equation}
\begin{split}
\frac{d \Gamma (K^+ \to \pi^+ \pi_0 \phi)}{dx_+ dx_0} &= \frac{|\Xi_A|^2}{f_\pi^2\Lambda^2}\frac{m_K^5}{512 \pi^3} (x_+-x_0)^2 \, , \\
\frac{d \Gamma (K_L \to \pi_0 \pi_0 \phi)}{dx_{0,a} dx_{0,b}} &= \frac{|\text{Re }\Xi_A|^2}{f_\pi^2\Lambda^2}\frac{m_K^5}{256 \pi^3} \left[\frac{2-x_{0,a}- x_{0,b}}{6} + \frac{m_\phi^2}{m_K^2} \left(2\sqrt{2}- \frac{1}{3}\right)\right]^2 \, , \\
\end{split}
\end{equation}
where we have defined $x_i = 2 p_K\cdot k_i/m_K^2$ with
$p_K$ being the kaon momentum and $k_i$ is the momentum belonging to one of the three final-state particles. 
Neglecting the mass difference between charged and neutral pions, the decay rate for these processes can be found by integrating the differential expressions between 
\begin{equation}
\label{eq:x2Boundaries}
    x_j^{\text{max, min}} = \frac{2 - 2 r - 3 x_i + r x_i + x_i^2 + 4 \varepsilon  - 2 x_i \varepsilon  \pm 
   \sqrt{x_i^2 - 4 \varepsilon} \, \sqrt{
     1 - 2 r + r^2 - 2 x_i + 2 r x_i + x_i^2 - 4 r \varepsilon}}{2 (1 - x_i + \varepsilon)} 
\end{equation}
and 
\begin{equation}
\label{eq:x1Boundaries}
    x_i^{\text{max}} = 1- r -2 \sqrt{\varepsilon r} \, , \qquad\qquad
    x_i^{\text{min}} = 2 \sqrt{\varepsilon} \,,
\end{equation}
where $r = m_\phi^2/m_K^2$ and $\varepsilon = m_\pi^2/m_K^2$ and the indices $i,j$ refer to pionic final-states only.
In the limit $m_K \gg m_\phi, m_\pi$, we find the following simplified expressions for the total decay rates: 
\begin{equation}
 \Gamma (K^+ \to \pi^+ \pi_0 \phi) \simeq \frac{|\Xi_A|^2}{f_\pi^2\Lambda^2}\frac{m_K^5}{6144 \pi^3} \, , \qquad
\Gamma (K_L \to \pi_0 \pi_0 \phi) \simeq \frac{|\text{Re }\Xi_A|^2}{f_\pi^2\Lambda^2}\frac{m_K^5}{6144 \pi^3} \frac{1}{6} \, . 
\end{equation}
We can then employ the E787 result $\text{BR}(K^+ \to \pi_0 \pi^+ \phi) < 3.8 \times 10^{-5}$ \cite{E787:2000iwe} and the E391 one, $\text{BR}(K_L \to \pi_0 \pi_0 \phi) < 0.7 \times 10^{-6}$ \cite{E391a:2011aa}, to put bounds on the coupling $\Xi_A$ appearing in the previous expressions: 
\begin{equation}
\begin{split}
|\Xi_A| \lesssim 1.1 \times 10^{-5} \frac{\Lambda}{\text{TeV}}\qquad \text{ and } \qquad 
|\text{Re }\Xi_A| \lesssim 1.7 \times 10^{-6} \frac{\Lambda}{\text{TeV}}\,.
\end{split}
\end{equation}

\section{Conclusions}
\label{sec:concl}
ALP theories are among the best motivated extensions beyond the SM, as they can naturally solve long-standing problems in particle physics such as the strong CP problem as well as the flavour and hierarchy problems while providing at the same time a viable dark matter candidate. 
ALP interactions with SM fermions and gauge bosons can be tested through a variety of
cosmological and astrophysical observables as well as through direct searches at collider.
Moreover, low-energy FCNC processes represent other powerful indirect probes of ALPs.
In this work, we have extended a previous analysis~\cite{DiLuzio:2020oah} - where CP-violating 
ALP scenarios have been discussed for the first time in the most general terms - assuming ALP 
masses below the GeV scale, where QCD cannot be treated perturbatively. 

Starting with a CP-violating effective Lagrangian containing light 
quarks and gluons, we have carefully discussed its matching onto a chiral 
effective Lagrangian, described in terms
of mesons and baryons, identifying the correspondence of the high-energy 
Jarlskog invariants with those emerging in the chiral theory.
This work provides the necessary tools for phenomenological analyses related 
to a light CP-violating ALP, connecting low-energy observables, such as EDMs 
and flavour processes entailing mesons and baryons, with the couplings of the 
underlying UV complete theory.
An extensive analysis of the phenomenological
implications of our work is worthwhile and it is postponed to a future study.

\section*{Acknowledgements}
 
We thank Gioacchino Piazza for useful comments on the manuscript.
 We thank Jorge Alda and Stefano Rigolin for useful comments and for having pointed out some typos in the previous versions of this paper.
This work received funding from the European Union's Horizon 2020 research and innovation programme under the 
Marie Sk\l{}odowska-Curie grant agreement n.~101086085 -- ASYMMETRY
and by the INFN Iniziative Specifica APINE. 
The work of LDL and PP 
is supported
by the European Union -- Next Generation EU and
by the Italian Ministry of University and Research (MUR) 
via the PRIN 2022 project n.~2022K4B58X -- AxionOrigins.
The work of LDL is funded by the European Union -- NextGenerationEU and by the University of Padua under the 2021 STARS Grants@Unipd programme (Acronym and title of the project: CPV-Axion -- Discovering the CP-violating axion).


\begin{small}

\bibliographystyle{utphys}
\bibliography{bibliography.bib}

\end{small}


\end{document}